\documentclass[journal,twoside,web]{ieeecolor}
\pdfoutput=1
\usepackage{etoolbox}
\makeatletter
\@ifundefined{color@begingroup}%
{\let\color@begingroup\relax
\let\color@endgroup\relax}{}%
\def\fix@ieeecolor@hbox#1{%
\hbox{\color@begingroup#1\color@endgroup}}
\patchcmd\@makecaption{\hbox}{\fix@ieeecolor@hbox}{}{\FAILED}
\patchcmd\@makecaption{\hbox}{\fix@ieeecolor@hbox}{}{\FAILED}

\usepackage{xcolor}
\usepackage{generic}
\usepackage{cite}
\usepackage{amsmath,amssymb}
\usepackage{algorithmic}
\usepackage{graphicx}
\usepackage{algorithm,algorithmic}
\usepackage{hyperref}
\hypersetup{hidelinks=true}
\usepackage{textcomp}
\usepackage{comment}
\usepackage{subfigure}
\usepackage{multirow}
\usepackage{makecell}
\usepackage{booktabs}
\usepackage{color}
\usepackage{ragged2e}
\usepackage{pifont}
\usepackage{bm}
\usepackage{tikz}
\usepackage{marvosym}
\usepackage{epstopdf}
\usepackage{marginnote}
\usepackage{threeparttable}
\usepackage{amsfonts}

\usepackage{verbatim}

\usepackage{hyperref}
\usepackage{url}
\usepackage{cite}
\hypersetup{colorlinks,
     citecolor=green,
     filecolor=green,
}

\definecolor{red}{rgb}{0.8,0,0}
\definecolor{blue}{rgb}{0,0,0.8}
\definecolor{green}{rgb}{0,0.4,0}
\newcommand{\change}[2]{}
\newcommand{\lchange}[2]{}

\newcommand{\changed}[3]{#3}

\newsavebox\CBox

\usepackage[T1]{fontenc}
\usepackage{listings}
\def\BibTeX{{\rm B\kern-.05em{\sc i\kern-.025em b}\kern-.08em
    T\kern-.1667em\lower.7ex\hbox{E}\kern-.125emX}}
\begin{document}
\clearpage
\twocolumn
\pagenumbering{arabic}
\setcounter{page}{1}
\setcounter{figure}{0}
\setcounter{table}{0}

\title{SLoRD: Structural Low-Rank Descriptors for Shape Consistency in Vertebrae Segmentation}
\author{Xin You, Yixin Lou, Minghui Zhang, Jie Yang, \IEEEmembership{Senior Member, IEEE}, Yun Gu, \IEEEmembership{Member, IEEE}
\thanks{This work is supported in part by National Natural Science Foundation of China under Grant 62373243, also in part by the Open Funding of Zhejiang Laboratory under Grant 2021KH0AB03.}
\thanks{X. You, Y. Lou, M. Zhang, J. Yang, and Y. Gu are with the Institute of Image
Processing and Pattern Recognition, Shanghai Jiao Tong University,
Shanghai 200240, China (Email: \{sjtu\_youxin, loulou\_0727,minghuizhang,jieyang, geron762 \}$@$sjtu.edu.cn).}
}

\maketitle

\begin{abstract}
Automatic and precise multi-class vertebrae segmentation from CT images is crucial for various clinical applications. However, due to 
similar appearances between adjacent vertebrae and the existence of various pathologies, existing single-stage and
multi-stage methods suffer from imprecise vertebrae segmentation. Essentially, these methods fail to explicitly impose both contour precision and intra-vertebrae voxel consistency constraints synchronously, resulting in the intra-vertebrae segmentation inconsistency, which refers to multiple label predictions inside a singular vertebra. \changed{M1.2}{\link{R1.2}}{In this work, we intend to label complete binary masks with sequential indices to address that challenge}. Specifically, a contour generation network is proposed based on Structural Low-Rank Descriptors for shape consistency, termed SLoRD. For a structural representation of vertebral contours, we adopt the spherical coordinate system and devise the spherical centroid to calculate contour descriptors. Due to vertebrae's similar appearances, basic contour descriptors can be acquired offline to restore original contours. Therefore, SLoRD leverages these contour priors and explicit shape constraints to facilitate regressed contour points close to vertebral surfaces. Quantitative and qualitative evaluations on VerSe 2019 and 2020 demonstrate the superior performance of our framework over other single-stage and multi-stage state-of-the-art (SOTA) methods. Further, SLoRD is a plug-and-play framework to refine the segmentation inconsistency existing in coarse predictions from other approaches. Source codes are available \url{https://github.com/AlexYouXin/SLoRD-VerSe}.
\end{abstract}


\begin{IEEEkeywords}
Vertebrae segmentation, Inconsistency, Contour descriptors, Refinement.
\end{IEEEkeywords}

\section{Introduction}
\begin{figure}[!t]
\centerline{\includegraphics[width=1.0\linewidth]{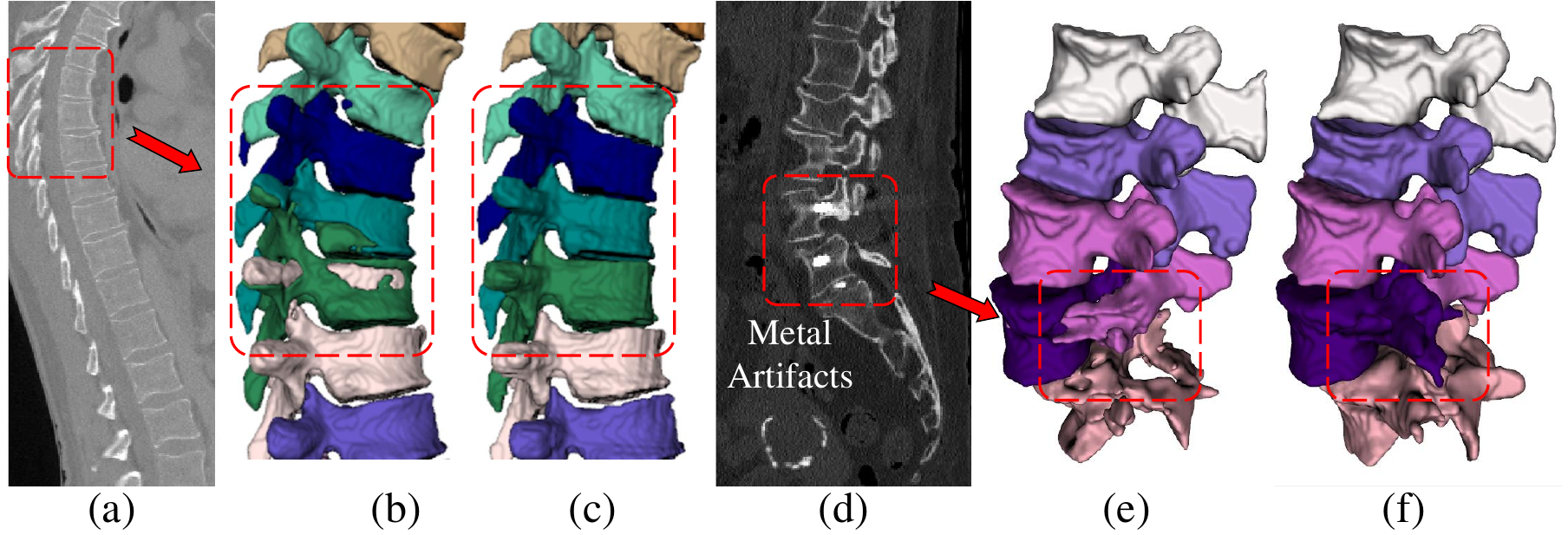}}
\caption{(a) CT case 1. (b-c) Coarse prediction by 3D UNet \& refined outcome by 3D UNet + SLoRD on case 1. (d) CT case 2 with metal artifacts. (e-f) Coarse prediction by the multi-stage pipeline Spine-Transformer \cite{tao2022spine} \& our refined prediction on case 2.}
\label{fig1}
\end{figure}

Automatic multi-class vertebrae segmentation from CT images with arbitrary field-of-views (FOVs) plays a significant role in spinal disease treatment, including preoperative diagnosis, surgical guidance, and postoperative assessment \cite{lessmann2019iterative, burns2016automated, knez2016computer}. Moreover, the shape analysis of vertebrae serve as an important anatomical reference for other anatomies in clinical practices \cite{frost2019materials, naegel2007using}. According to the anatomical prior of vertebral structures, it is emerging to preserve the consistency of label predictions inside a singular vertebra~\cite{lessmann2019iterative}. 
However, due to similar appearances between adjacent vertebrae and the existence of various pathologies \cite{sekuboyina2021verse, tao2022spine, mao2024semantics, you2024learning}, single-stage and multi-stage methods are prone to generate inconsistent mask predictions, which contain inconsistent label predictions for a singular vertebra as illustrated in Fig.\ref{fig1}. That phenomenon can be recognized as intra-vertebrae label inconsistency \cite{you2023verteformer}, which poses an enormous obstacle for the task of precise vertebrae segmentation.

Specifically, earlier methods solved vertebrae segmentation from volumetric CT images by employing statistical models \cite{klinder2009automated}, active shape models \cite{hammernik2015vertebrae}, level sets \cite{lim2014robust}, Markov random field \cite{kadoury2011automatic, kadoury2013spine}, graph cuts \cite{aslan20103d} and machine learning theories \cite{chu2015fully, kelm2013spine}. However, those approaches fail to perform well on unseen data due to insufficient generalization ability. Current methods based on deep learning techniques demonstrate a more promising performance for vertebrae image segmentation even with arbitrary FOVs. One category of existing works \cite{cciccek20163d, isensee2021nnu, you2023verteformer} adopted \textbf{single-stage} convolutional or Transformer-based networks for automatic vertebrae segmentation and identification. However, these models suffer from drawbacks of similar appearances between adjacent vertebrae and the patch-based inference paradigm \cite{lessmann2019iterative}, resulting in intra-vertebrae label inconsistency as revealed in Fig.\ref{fig1}(a \& b). The other methods \cite{meng2023vertebrae, masuzawa2020automatic, tao2022spine, payer2020coarse} designed a \textbf{multi-stage} pipeline, which can boost the segmentation performance by reducing the disturbance from neighboring contexts based on the localize-then-segment concept. However, a biased local patch from the detection network due to the inaccurate center regression will enforce inconsistency inside vertebrae, particularly for pathological vertebrae with abnormal shapes or metal implant artifacts as illustrated by Fig.\ref{fig1}(d \& e). Also, this limitation can be referred to according to the qualitative analysis of experimental results in \cite{tao2022spine}.

Essentially, all the above methods fail to explicitly impose both contour precision and intra-vertebrae voxel consistency constraints synchronously, resulting in segmentation inconsistency. To resolve this limitation, \changed{M1.2}{\link{R1.2}}{we attempt to label complete binary masks with sequential indices to achieve multi-label vertebrae segmentation.} Due to the connection between superior and inferior articular processes \cite{burns2016automated}, the main challenge is transformed to how to divide joint binary vertebral masks while preserving original contours to the greatest extent \cite{gafencu2024shape}. 

\setcounter{figure}{1}
\renewcommand{\thefigure}{\arabic{figure}}
\begin{figure}[!t]
\centerline{\includegraphics[width=0.95\linewidth]{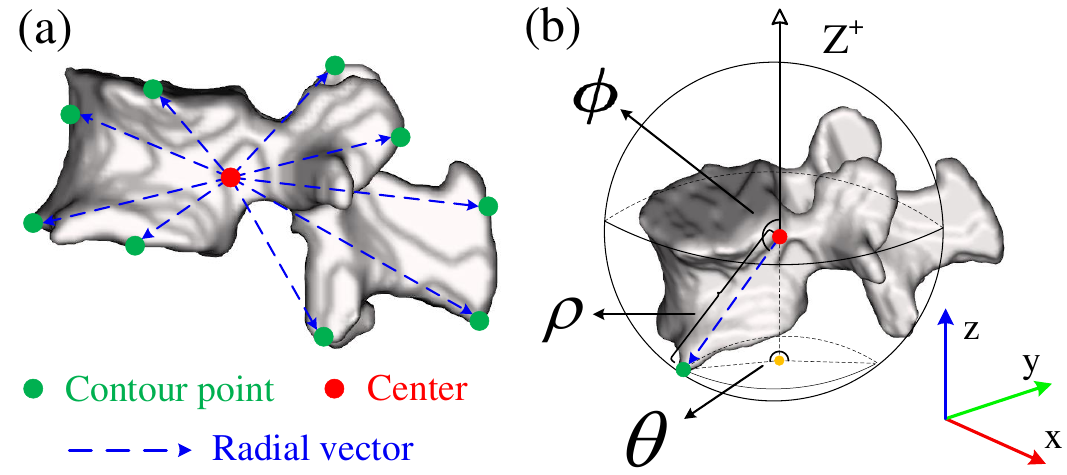}}
\caption{Spherical coordinate system for contour representations of 3D vertebrae. (a). red point: the spherical center, green point: the contour point, blue line: the radial vector. (b). x, y, and z axis adopt the orientations shown in the figure. $\rho$: the radial distance away from the spherical center, $\phi$: the azimuth angle between the radial vector and $z^{+}$, $\theta$: the polar angle between $x^{+}$ and the projected radial vector in the XOY plane.}
\label{spherical_system}
\end{figure}

Thus, we devise an effective contour generation network based on Structural Low-Rank Descriptors for shape consistency, termed SLoRD. For the contour representation, the complete point cloud of vertebral boundaries shows an unorganized arrangement with various point numbers. Intuited by \cite{xie2020polarmask}, we propose a novel way of contour representations based on the spherical coordinate system as shown in Fig.\ref{spherical_system}. Under that circumstance, structural contour descriptors are efficiently calculated by recording the radial distance. Additionally, spherical centroids are proposed to enhance the precision of contour representations, serving as the spherical center. Then, vertebrae show similar 3D shapes according to anatomical structures, lumbar and thoracic vertebrae in particular. Hence, low-rank contour descriptors can be extracted from the training dataset in an offline manner. Via a linear combination between low-rank basic descriptors in the contour space, instance masks of vertebrae can be approximately restored. Further, vertebral labels provided by coarse predictions facilitate the consistent generation of semantic masks. Ultimately, the whole pipeline is incorporated into SLoRD to boost the framework's generalization abilities on unseen datasets. And the explicit shape constraint and basic contour descriptors are exerted to promote the regression precision of contours. SLoRD serves as a plug-and-play framework, which can repair the segmentation inconsistency from coarse predictions.

\changed{M7.2}{\link{R7.2}}{Our contributions are summarized as follows: (1) To achieve intra-vertebrae voxel consistency and contour precision simultaneously, we first propose the contour-based generation network named SLoRD, to address the multi-class vertebrae segmentation from CT images with different FOVs. (2) For a structural and efficient contour representation, the spherical coordinate system is adopted to describe the complete contour point cloud. Besides, spherical centroids are devised as the spherical center to boost the precision of contour representations, based on the anatomical characteristics of vertebrae. (3) Profiting from low-rank basic descriptors in the contour space, regressed contour points will be facilitated close to vertebral surfaces with shape and contour constraints. (4) Extensive experiments on VerSe 2019 and 2020 datasets demonstrate the superior performance of our methods compared with other single-stage and multi-stage approaches, with better segmentation consistency and boundary precision. Moreover, the SLoRD framework serves as a plug-and-play framework to repair intra-vertebrae segmentation inconsistency significantly.}

\begin{figure*}[!t]
\centerline{\includegraphics[width=0.95\linewidth]{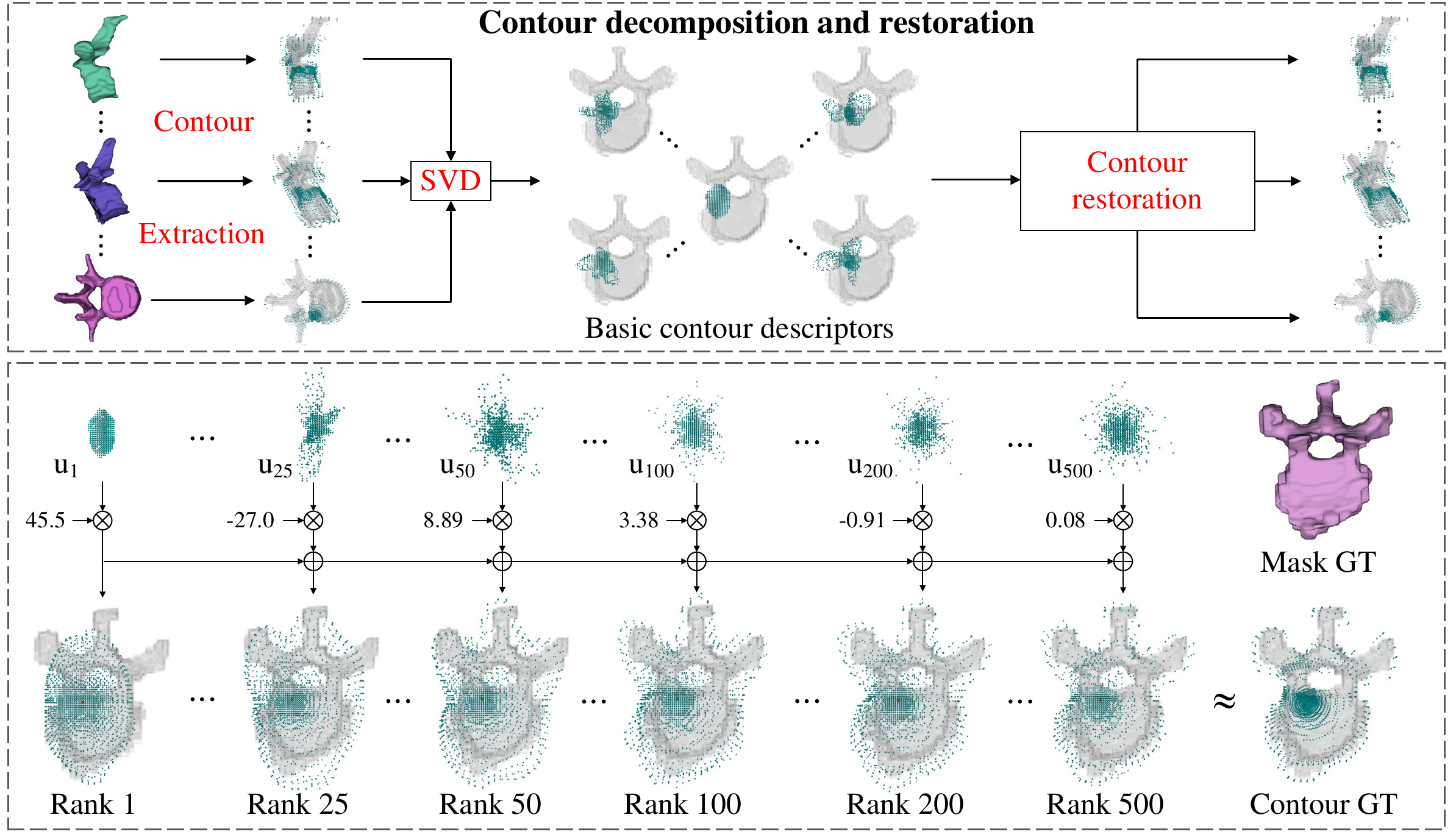}}
\caption{Upper: The flow of contour decomposition and restoration. Lower: The linear combination process between basic contour descriptors. The first row indicates basic descriptors of specific ranks. The second row refers to weighted visualizations.}
\label{fig2}
\end{figure*}

\section{Related work}
While the shape of individual vertebrae varies considerably along the spine, adjacent vertebrae still show fairly similar appearances, making them difficult to distinguish \cite{pang2020spineparsenet, you2022eg}. Consequently, the spine presents unique challenges for quantitative image analysis compared to other anatomical structures. Additionally, factors such as pathological appearances, fractures, and metal implant artifacts \cite{you2023verteformer} will introduce further challenges. Earlier studies \cite{sekuboyina2017localisation, janssens2018fully, al2018fully, al2018spnet}, cannot apply to the segmentation task of CT scans displaying arbitrary FOVs with various label distributions \cite{wu2023multi, meng2023vertebrae}. To resolve these challenges, researchers resort to leveraging anatomical prior knowledge. Relevant research could be roughly categorized into two groups: single-stage and multi-stage frameworks for multi-class vertebrae segmentation.

For single-stage pipelines, Zheng et al. employed nnUNet \cite{isensee2021nnu} to perform multi-class vertebrae segmentation. Chang et al. \cite{chang2020multi} adopted stacked graph convolutional networks by featuring a pre-defined adjacency matrix to encode structural information. However, such approaches fail to model global dependency due to the intrinsic property of convolution \cite{dosovitskiy2020image}. Thus, Verteformer \cite{you2023verteformer} was devised by fusing local details and global contexts with two parallel Transformers. And an edge detection block was proposed to refine intra-vertebrae boundaries. Nevertheless, it is still plagued by similar shapes between adjacent vertebrae and the patch-based inference paradigm, leading to intra-vertebrae segmentation inconsistency. Lessmann et al. \cite{lessmann2019iterative} used fully convolutional networks by iteratively analyzing image patches for automatic vertebrae segmentation and identification. However, this method relies on the perfect localization of the topmost vertebra using a sliding window fashion, thus is prone to sequential poor segmentation without a precise detection of the topmost vertebra.


In contrast, multi-stage approaches reveal more promising segmentation performance owing to the additional guidance, such as the heatmap and box-based localization of vertebrae. Sekuboyina et al. \cite{sekuboyina2021verse} proposed an interactive spine processing framework, consisting of spine detection by a light-weight network, vertebra labeling based on Btrfly Net \cite{sekuboyina2018btrfly}, and vertebral segmentation by 3D UNet \cite{cciccek20163d}. Payer et al. \cite{payer2020coarse} implemented a three-step approach to first localize the spine, then simultaneously locate and identify each vertebra, and finally segment each vertebra. Besides, Tao et al. \cite{tao2022spine} introduced a Transformer to handle the labeling task, then employed UNet with the original CT and an auxiliary heatmap as inputs to segment all vertebrae. Chen et al. \cite{chen2020deep} devised a deep reasoning module leveraging anatomical prior knowledge to achieve anatomically correct results regarding the sequence of vertebrae. However, the multi-stage pipelines are prone to accumulating errors. An imprecise detected patch with inaccurate
center regression will degrade the segmentation performance, particularly for pathological vertebrae with abnormal shapes
or metal implants. As a result, the poor contour delineation will enforce inconsistency inside vertebrae.

\section{Methodology}
To simultaneously impose the voxel consistency and contour precision regularizations, we propose SLoRD by labeling sequential and complete binary masks to achieve multi-class vetebrae segmentation. Specifically, a spherical-based representation mode and spherical centroid are devised for structural and efficient contour representations as described in Section \ref{contour descriptors} and \ref{Spherical centroid}. Section \ref{SLoRD} provides the details of SLoRD, which generates precise contour points based on explicit basic descriptors and contour regularization. The pipeline of the inference stage is illustrated in Section \ref{inference}.
 
\subsection{Spherical-based Contour Descriptors}
\label{contour descriptors}
\changed{M3.2.1}{\link{R3.2}}{Although the number of contour points varies across different vertebrae, a fixed number of sampling points is required for contour representations due to the predefined network. However, it is quite tricky to sample contour points uniformly while maintaining the potential to restore the original contour. Thus, we incorporate the idea from PolarMask \cite{xie2020polarmask, park2022eigencontours}, in which each contour point is encoded with the polar coordinate.}
\begin{figure*}[!t]
\centerline{\includegraphics[width=0.95\linewidth]{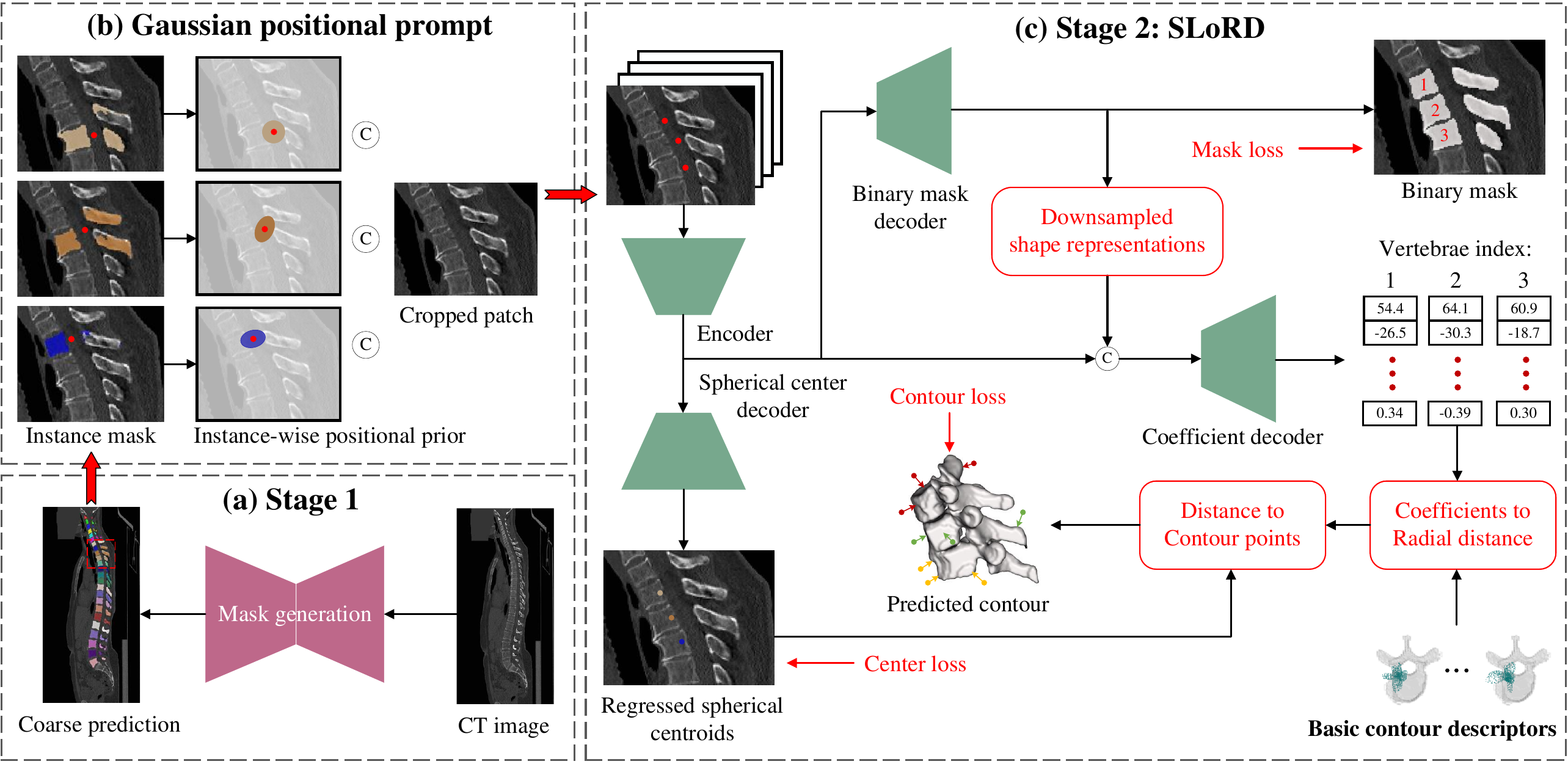}}
\changed{M5.4}{\link{R5.4}}{\caption{The two-stage framework. (a) The first stage adopts arbitrary segmentation networks to output inconsistent predictions, which will be refined with intra-vertebrae segmentation consistency by SLoRD in a sliding-window fashion. (b) Gaussian positional prompts are acquired based on the position of coarse spherical centers and the size of instance masks. Instance-wise positional priors are beneficial to promote the spherical centroid localization by SLoRD. (c) SLoRD consists of the spherical center decoder, the coefficient decoder, and the binary mask decoder. Specifically, the center decoder will regress spherical centroids with better precision. The coefficient decoder will generate linear coefficients, which are transformed into contour points under the interaction with low-rank contour descriptors. The mask decoder is aimed for precise shape representations of vertebrae, which boost the precision of generated contours. SLoRD aims to regress contour points close to vertebral boundaries via explicit contour and shape regularizations.}}
\label{fig3}
\end{figure*}
Given the difficulty in extending the polar system to contour representations of 3D objects like vertebrae, we formulate contours with the spherical coordinate system as shown in Fig.\ref{spherical_system}. \changed{M4.5}{\link{R4.5}}{Conditioning the inner centroid as the spherical center, we describe the boundary using spherical coordinates $(\rho, \theta, \phi)$.} Here $\rho$ represents the length of the radial vector from the spherical center to the contour point, $\phi$ is the azimuth angle between the radial vector and $z^{+}$, $\theta$ means the polar angle between $x^{+}$ and the projected radial vector in the XOY plane. Angular coordinates $\theta_{i}$ and $\phi_{j}$ are sampled uniformly with a sampling interval $s$, with $I$ equal to $360/s$, $J$ equal to $(180/s \mbox{+} 1)$.
Thus, given the sampling angle grid, only radial coordinates are recorded to represent the contour:
\setcounter{equation}{0}
\renewcommand{\theequation}{\arabic{equation}}
\begin{eqnarray}
  & \bm{\rho} = [\rho_{_\mathrm{11}}, ... , \rho_{_\mathrm{1J}}, ... , \rho_{_\mathrm{I1}}, ..., \rho_{_\mathrm{IJ}}]^{T} \\
    & \rho_{_\mathrm{ij}} = \rho(\theta_{i}, \phi_{j}), \, \, i=1,...,I; \; j=1,...,J
\end{eqnarray}
\changed{M4.5}{\link{R4.5}}{here $\bm{\rho}$ refers to the contour descriptor, with the dimension $N$ equal to $I \times J$. $\rho_{ij}$ means the radial length from the spherical center to the specific contour point with the polar angle and azimuth angle equal to $\theta_{i}$ and $\phi_{i}$.} Considering vertebral anatomies with the non-convex shape property, the spherical system chooses the maximum distance at the same angle to strike a balance between precision and computational cost. Since the shapes of most vertebrae are well structured and resembling to each other, low-rank contour descriptors can be exploited in a data-driven manner via Singular Value Decomposition (SVD) \cite{park2022eigencontours}. Detailedly, a non-convex contour matrix $\mathcal{M} = [\bm{\rho}_{1}, \cdot \cdot \cdot  , \bm{\rho}_{L}]$ is constructed from L vertebrae in the training data. Then we perform SVD on matrix $\mathcal{M}$:
\begin{eqnarray}
  & \mathcal{M} = \bm{U} \, \Sigma \, \bm{V}^{T}
\end{eqnarray}
where $\bm{U}^{N \times N}=[\bm{u}_{1}, \cdot \cdot \cdot , \bm{u}_{N}]$ and $\bm{V}^{L \times L}=[\bm{v}_{1}, \cdot \cdot \cdot , \bm{v}_{L}]$ are orthogonal matrixes and $\Sigma^{N \times L}$ is a rectangular diagonal matrix, consisting of singular values $[\sigma_{1}, \cdot \cdot \cdot , \sigma_{r}]$ with r equal to $\min(N, L)$. The best rank-k approximation of $\mathcal{M}$ can be calculated as follows:

\begin{equation}
  \begin{aligned}
     \mathcal{M}(k) &=  \sigma_{1} \bm{u}_{1} \bm{v}_{1}^{T} + \cdot \cdot \cdot + \sigma_{k} \bm{u}_{k} \bm{v}_{k}^{T}  \\
  &= [\bm{u}_{1}, \cdot \cdot \cdot , \bm{u}_{k}] \, \mathcal{C} = \bm{U}(k) \, \mathcal{C} \approx \mathcal{M} \\ 
 & \hspace{-9mm} \mathcal{C} = [\sigma_{1}\bm{v}_{1}, \cdot \cdot \cdot , \sigma_{k}\bm{v}_{k}]^{T} = \bm{U}(k)^{T} \mathcal{M} \\
  \end{aligned}
\label{rank-k}
\end{equation}
\changed{M3.2.2}{\link{R3.2}}{Here $\bm{U}(k)^{N \times k}$ refers to k basic contour descriptors, which can span into the contour space and be acquired from the training data in advance. $\mathcal{C}^{k \times L}$ means combinational coefficients, calculated by eigen descriptors and matrix $\mathcal{M}$. As long as $\mathcal{C}$ is correct, we can restore the original contours of vertebrae approximately as illustrated by Fig.\ref{fig2}.}

\changed{M4.4.1}{\link{R4.4}}{Another contour-based DeepSSM \cite{bhalodia2024deepssm}, as a shape model based on principal component analysis (PCA), aims to learn the functional mapping from images to low-dimensional latent codes $\mathcal{C}_{l}$. Then the model achieves discrete contour representations $\mathcal{R}_{c}$ by applying the linear projection on these codes with the PCA basis $\mu$ and mean shape $\nu$. This process can be described as follows:
\begin{eqnarray}
  & \mathcal{R}_{c} = \mu \,\mathcal{C}_{l} + \nu
  \label{deepssm}
\end{eqnarray}
In contrast, SVD bears the best low-rank approximation properties, and extracted basic descriptors are more robust to fit various anatomical shapes than PCA \cite{wall2003singular}. Also, the sampling strategy in \cite{bhalodia2024deepssm} for the contour point cloud struggles to perform well for pathological vertebral structures with more diverse shapes and topology.}
\subsection{Spherical Centroid}
\label{Spherical centroid}
Since the azimuth angle $\phi$ and polar angle $\theta$ are both sampled at equal intervals, a spherical center with a biased location will result in sparsely sampled contour points for the boundary region far from the center. Consequently, that region suffers from a contour under-representation, which further affects the restoration of vertebral shapes. Hence, it is crucial to designate the spherical center adaptive to the specific shape information of vertebrae.

\changed{M4.6}{\link{R4.6}}{A simple and straightforward configuration is the vertebral centroid, averaged on both the vertebral body and spinous process. However, the centroid is acquired ignoring the optimization objective of enforcing contour points close to the boundary $\mathcal{B}$. Specifically, in the process of constructing the contour matrix $\mathcal{M}$, angles corresponding to boundary points are represented as floating-point numbers. To improve computational and storage efficiency, these angles are approximated to integers. Consequently, sampling contour points based on an integer-indexed angle grid inevitably introduces approximation errors, causing the sampled contour points to deviate from the true boundary surface, which consequently affects the accuracy of vertebral shape reconstruction. Thus, to mitigate this deviation, a distance-based center should be marked by minimizing the average Euclidean distance between this center and the whole boundary surface.} We formulate the distance-based center as spherical centroid $C_{s}$ based on Eq \eqref{center point}:
\begin{eqnarray}
\begin{aligned}
  &  \mathcal{B} = \mathcal{F}(M) \setminus M \\
  & C_{s} = \mathop{\arg\min} \limits_{c \in M} (\tfrac{1}{|\mathcal{B}|} \displaystyle \sum_{b \in \mathcal{B}} d \,(c, b) + \lambda \cdot d \,(c, \delta)),  \\ 
  & \mathrm{ s.t. } \ \delta = \mathop{\arg\max} \limits_{j \in \mathcal{B}} j_{y} 
\end{aligned}
  \label{center point}
\end{eqnarray}
where $\mathcal{F}(\cdot)$ represents the morphological dilation operation, $M$ refers to the segmentation mask of each vertebral volume, the boundary map $\mathcal{B}$ is the difference set between the dilated mask $\mathcal{F}(M)$ and the mask itself $M$. \changed{M3.3.2}{\link{R3.3}}{Besides, $d(\cdot)$ means the $L_{2}$ norm operation in Euclidean space.} The first item in Eq \eqref{center point} refers to the average distance from each boundary point $b$ to the center $c \in \mathcal{P}$. \changed{M4.6}{\link{R4.6}}{Concerning the non-convex shape prior of the spinous process, it will be more difficult to approximate finer contours of this region. Therefore, we attempt to push the spherical centroid closer to it. As illustrated by Fig.\ref{fig2}, that will promote a denser sampling of contour points near the spinous process, then enhance the shape representations of the spinous process. That is why the second regularization term is incorporated,} in which $\delta$ is the boundary point $j \in \mathcal{B}$ with the largest y-dimensional index $j_{y}$. $\lambda$ is a scaling parameter, empirically set as $0.005$.

\subsection{Contour Generation with Explicit Contour Priors}
\label{SLoRD}
To precisely restore the contour with explicit contour descriptors, it is significant to accurately achieve the coefficient according to Eq \eqref{rank-k}. Besides, the contour generation requires the precise regression of spherical centers as described in Section \ref{Spherical centroid}. Thus, we devise the SLoRD framework by introducing the center decoder and coefficient decoder as shown in Fig.\ref{fig3}(c). 

\textbf{Center decoder:} Deep features from the encoder are transformed into regressed spherical centers via combinative operations of convolution, batch normalization, average pooling. The pooling process projects the feature dimension as 3, representing coordinates of the regressed center $\hat{C}=(\hat{c}_{x}, \hat{c}_{y}, \hat{c}_{z})$. The ground truth of center $C_{s}$ can be attained by Eq \eqref{center point}. And we employ the $L_{2}$ norm to regularize regressed centers $\hat{C}$, with the center loss $\mathcal{L}_{center}$ equal to $|| \, C_{s} - \hat{C} \,||^{2}$. Concerning the limitation that some methods wrongly detect centers of specific pathological vertebrae, we aim to introduce Gaussian positional prompts as additional inputs for positional guidance. As shown in Fig.\ref{fig3}(b), the positional prompt is defined as a Gaussian distribution $\mathcal{G}$ based on the following formula:
\begin{eqnarray}
& \mathcal{G} (\bm{\mu}; \bm{\sigma}) = \mathcal{G} (\mu_{x}; \sigma_{x}) \cdot \mathcal{G} (\mu_{y}; \sigma_{y}) \cdot \mathcal{G} (\mu_{z}; \sigma_{z})    \\
& \bm{\sigma} = \mathop{\max}(\bm{m}, \bm{\bar m}) \; / \; 4, \ \bm{\bar m} = \Large \tfrac{1}{L} \displaystyle \sum \bm{m}
  \label{gaussian priors}
\end{eqnarray}
Specifically, the mean value $\bm{\mu}=(\mu_{x}, \mu_{y}, \mu_{z})$ is set as the centroid based on the coarse prediction. For defining the variance $\bm{\sigma}=(\sigma_{x}, \sigma_{y}, \sigma_{z})$, we aim to cover the whole shape distribution using the given $\mathcal{G}$. Thus, as suggested by the two-sigma rule in mathematical statistics, the length of $4\bm{\sigma}$ should encompass the 3-dimensional length $\bm{m}=(m_{x}, m_{y}, m_{z})$ of each instance mask. Further, a threshold $\bm{\bar m}$ is introduced to ensure a wide enough region with attention weights decaying from the center $\bm{\mu}$ to the vertebral boundary. \changed{M4.7}{\link{R4.7}}{The center decoder will modify coarse centroids $\bm{\mu}$ as refined spherical centroids via the explicit constraint $\mathcal{L}_{center}$.}

\textbf{Coefficient decoder:} Similar to the center decoder, this part is also a combination of convolution, normalization and average pooling layers, with the projected dimension equal to the rank size. Here a direct loss on combinational coefficients like the previous work \cite{park2022eigencontours} cannot work in this task. We reckon that phenomenon results from the magnitude difference between different dimensions of coefficient vectors and the magnitude variance between coefficient vectors at the same dimension. Thus, we transform coefficients into the radial distance with data-driven contour priors, then into the contour point cloud \cite{weisstein2005spherical} centered at spherical centroids. After that, an explicit contour constraint $\mathcal{L}_{contour}$ is devised based on the average minimal distance from the contour point set $\mathcal{P}_{c}$ to the boundary interface $\mathcal{B}$ as illustrated by the following formula:
\changed{M6.6}{\link{R6.6}}{\begin{align}
   \mathcal{L}_{contour} &= \Large \tfrac{1}{|\mathcal{P}_{c}|} \displaystyle \sum_{p \in \mathcal{P}_{c}} \mathop{\min} \limits_{b \in \mathcal{B}} d(p, b) \\ 
 &= \Large \tfrac{1}{|\mathcal{P}_{c}|} \displaystyle \sum_{p \in \mathcal{P}_{c}} \mathop{\min} \limits_{b \in \mathcal{B}} \sqrt{\| p - b \|^{2}}
  \label{contour loss}
\end{align}
}
where $|\mathcal{P}_{c}|$ means the voxel number of $\mathcal{P}_{c}$. Besides, it might be confusing why a straightforward distance regularization is not conducted on the regressed point clouds. \changed{M3.2.1}{\link{R3.2}}{A comprehensible perspective is that SLoRD does not bear any perception ability to identify the sampled angle grid in the spherical coordinate system with no angular priors introduced. That is why basic contour descriptors will boost the contour representation of vertebral anatomies.}

\textbf{Loss term:} \changed{M3.5}{\link{R3.5}}{Guided by the recent work \cite{bhalodia2024deepssm, bhalodia2018deepssm}, $\tfrac{1}{3}$ uniformly sampled boundary points can approximately restore the shape information of vertebrae.} Thus, uniformly sampled boundary points from $\mathcal{B}$ is incorporated to calculate the contour loss. \changed{M3.3.1}{\link{R3.3}}{Besides, an additional binary mask supervision $\mathcal{L}_{mask}$ (cross-entropy \mbox{+} Dice loss) will exert implicit shape information. The binary mask decoder is configured with the structure of MedNeXt decoder \cite{roy2023mednext}, boosting the regression of vertebral contours.} Thus, the total loss is defined as follows:
\begin{eqnarray}
  & \mathcal{L} = \mathcal{L}_{mask} + \mathcal{L}_{center} + \mathcal{L}_{contour}
  \label{total loss}
\end{eqnarray}
Furthermore, segmentation inconsistency tends to appear in a sequence of vertebrae as highlighted in Fig.\ref{fig1}. Under this circumstance, the iterative refinement for one vertebra each time cannot well address this type of failure cases, hence we attempt to handle three sequential vertebrae once. Assuming that coarse predictions reveal satisfactory performance for the top two vertebrae, even with a moderate inconsistency, SLoRD can significantly boost segmentation results. 

\begin{table*}[!t]
  \begin{center}
  \caption{Comparisons with other single-stage, two-stage, and multi-stage methods on the public and hidden test datasets of VerSe 2019 (A-Dice: average Dice. M-Dice: median Dice. A-HD: average Hausdorff distance. M-HD: median Hausdorff distance).}
  \label{tab1}
    \resizebox{0.94\textwidth}{!}{
  \begin{tabular}{clcccccccc}  
  \hline
 \multirow{2}*{Stage} &  \multirow{2}*{Method} & \multicolumn{4}{c}{\textbf{Public test}} & \multicolumn{4}{c}{\textbf{Hidden test}}  \\  
  \cmidrule(r){3-6}    \cmidrule(r){7-10}
  & & A-Dice $\uparrow$ & M-Dice $\uparrow$ & A-HD $\downarrow$ & M-HD $\downarrow$ & A-Dice $\uparrow$ & M-Dice $\uparrow$ & A-HD $\downarrow$ & M-HD $\downarrow$ \\
  \hline  
  & 3D UNet \cite{cciccek20163d} & 81.13  & 94.71  & 16.36  & 13.81  & 81.28 & 87.54 & 16.35 & 14.73  \\
    & Lessmann N.\cite{lessmann2019iterative}  & 85.08 & 94.25 & 8.58 & 4.62 & 85.76 & 93.86 & 8.20 & 5.38 \\
  & nnUNet \cite{isensee2021nnu} & 85.81 &  92.55 & 12.50 & 10.25 & 86.59 & 92.62 & 12.78 & 12.00 \\
  Single-stage &  Verteformer \cite{you2023verteformer} & 86.39 & 91.22  & 11.14 & 10.28 & 86.54 & 90.74  & 10.55 & 10.51 \\ 
  & UNeXt\cite{valanarasu2022unext} & 83.80 & 85.20  & 18.63 & 13.51 & 83.36 & 88.39  & 12.19 & 11.77\\
  & Swin UNETR \cite{tang2022self} & 85.42 & 90.57  & 16.01 & 12.55 & 83.46 & 88.91  & 16.03 & 13.58 \\
  & MedNeXt \cite{roy2023mednext} & 88.51 & 93.96 & 10.59 & 8.47 & 88.91 & 94.40  & 11.35 & 9.81 \\
  \hline
  Multi-stage & Sekuboyina A. \cite{sekuboyina2021verse} & 83.06  & 90.93 & 12.11 & 7.56 & 83.18 & 92.79 & 9.94 & 7.22 \\ 
  Multi-stage & Payer C. \cite{payer2020coarse} & \underline{90.90}  & \underline{95.54} & \underline{6.35} & 4.62 & 89.80 & \underline{95.47} & 7.08 & \underline{4.45} \\
  Two-stage & SAM-Med2D \cite{cheng2023sam} & 89.37 & 93.49  & 8.95 & 6.77 & 89.26 & 94.34 & 8.85 & 5.69 \\
  Two-stage & Tao R. \cite{tao2022spine}  & 90.49 & 94.26  & 6.53 & \underline{4.31} & \underline{89.94}  & 94.01 & \underline{6.96} & 4.73 \\
    \hline

  Two-stage & Ours  & \textbf{92.17} & \textbf{95.62} & \textbf{5.91} & \textbf{4.06} & \textbf{91.33} & \textbf{96.03} & \textbf{5.97} & \textbf{4.27} \\
  \hline  
  \end{tabular}}
  \end{center}
\end{table*}

\begin{figure}[!t]
\centerline{\includegraphics[width=1.0\linewidth]{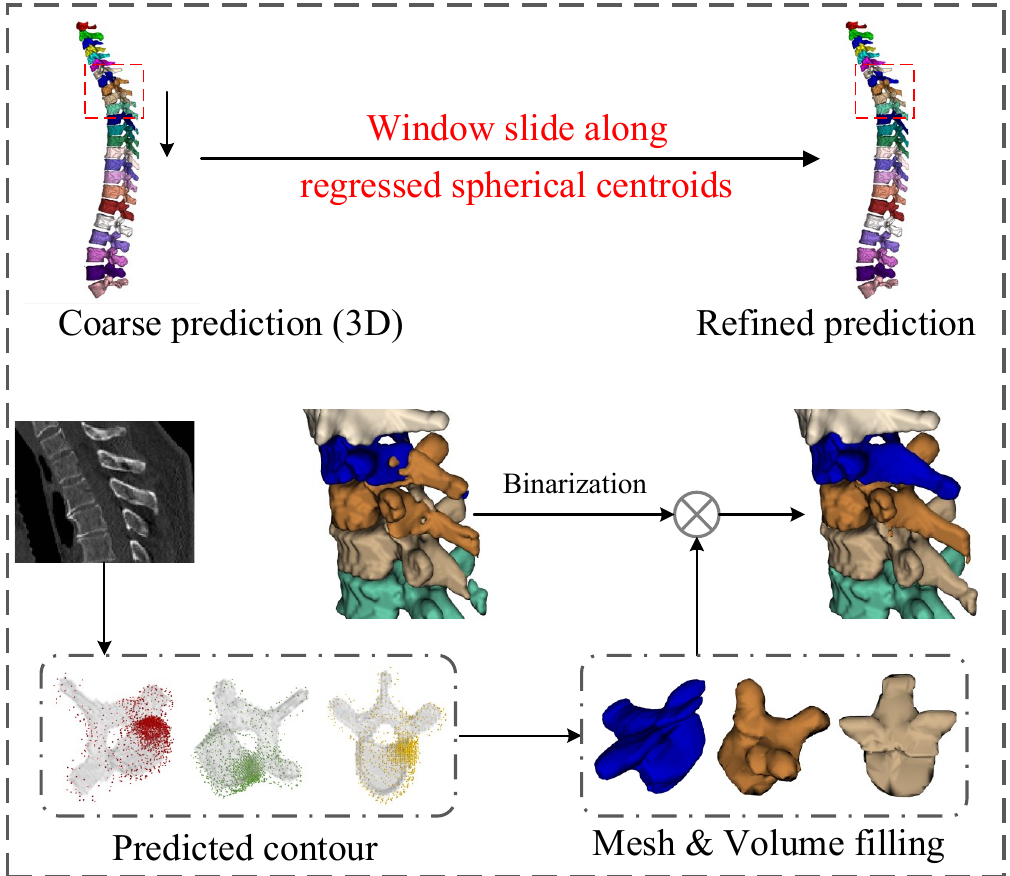}}
\caption{Iterative refinement for coarse masks in the inference stage.}
\label{inference_pipeline}
\end{figure}

\subsection{Inference}
\label{inference}
In the inference phase, the patch to be refined in SLoRD is cropped along each vertebral center from coarse predictions by arbitrary segmentation networks as shown in Fig.\ref{fig3}(a). The SLoRD framework will first refine three spherical centers corresponding to sequential vertebrae according to the prediction of the center decoder. Then the coefficient branch will output $3 \times N$ (N refers to the dimension of $\bm{\rho}$) contour points surrounding three vertebral surfaces with different labels as revealed in Fig.\ref{inference_pipeline}. After that, Marching Cube \cite{lorensen1998marching} is implemented to generate triangle meshes, which are further voxelized into a filled volume \cite{schroeder1998visualization}. \changed{M1.5}{\link{R1.5}}{Lastly, we perform the binarized attention operation $\mathcal{O}_{B}$ on the set of coarse masks $\mathcal{S}_{C}$ and labeled masks $\mathcal{S}_{L}$ produced by SLoRD. The mathematical formulation is as follows:
\begin{eqnarray}
    \mathcal{S}_{R} = \mathcal{O}_{B}(\mathcal{S}_{C}, \mathcal{S}_{L}) = \mathbb{1} (\mathcal{S}_{C}) \cdot \mathcal{S}_{L}    \\
    \mathbb{1} (i)=
\begin{cases}
 1, & \text{if }\; i > 0 \\
0, & \text{otherwise }
\end{cases}
\end{eqnarray}
we can acquire the set of refined segmentation masks $\mathcal{S}_{R}$, which reveals both good label consistency and shape restoration via the auxiliary connected component analysis.}

\begin{figure*}[!t]
\centerline{\includegraphics[width=0.98\linewidth]{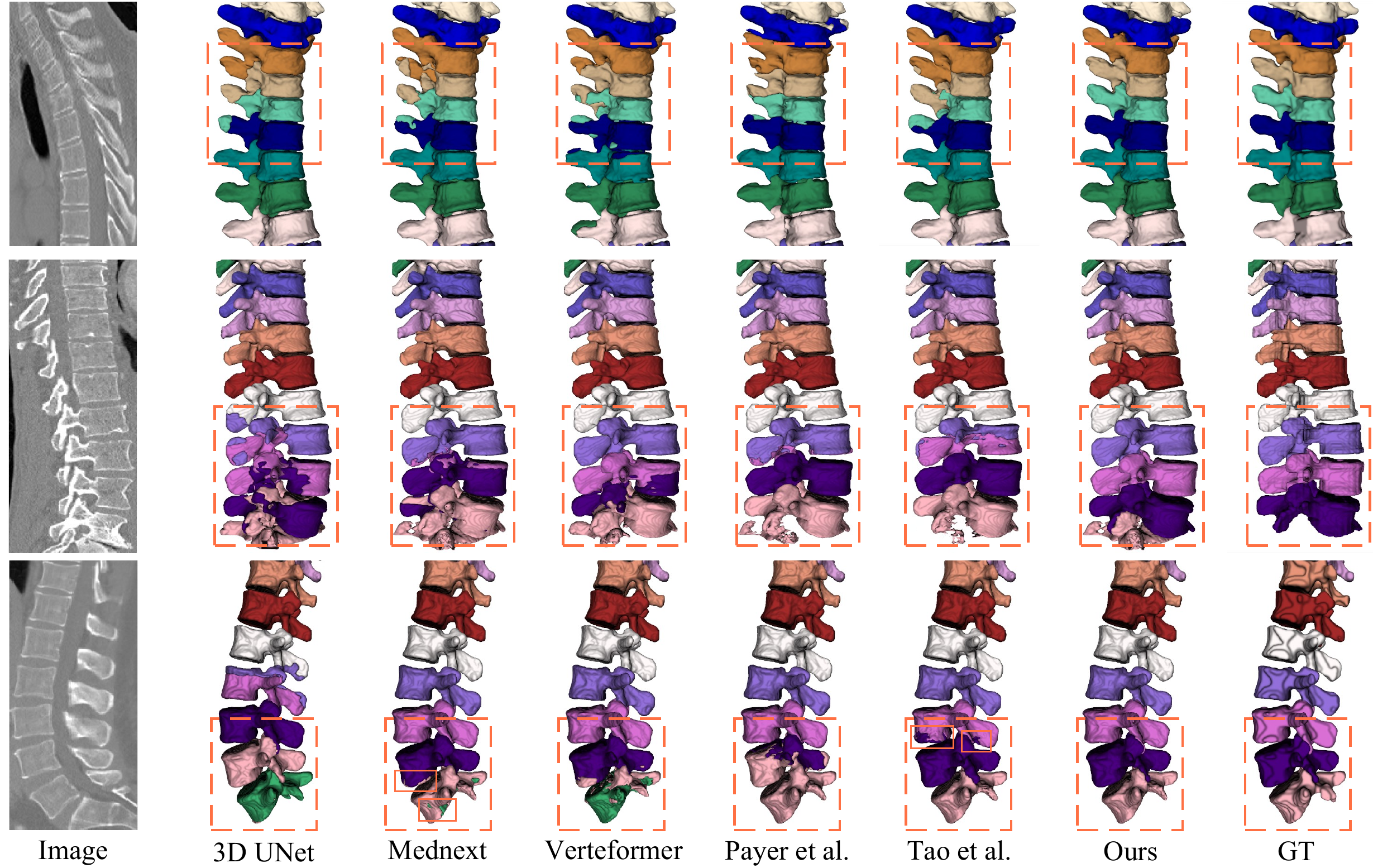}}
\caption{Qualitative visualizations of hard cases by single-stage, two-stage, and multi-stage models in the benchmark. The $1^{st}$ row: a case from VerSe 2019. The $2^{nd}$ and $3^{rd}$ row: two cases from VerSe 2020.}
\label{benchmark}
\end{figure*}

\section{Experiment}
\subsection{Experimental Settings}
\label{experimental settings}
\textbf{VerSe 2019 \& 2020.} 
To assess the efficacy of our method, we conduct experiments on the Large Scale Vertebrae Segmentation Challenge (VerSe 2019 \& 2020) \cite{sekuboyina2021verse}. Specifically, VerSe 2019 consists of 160 spinal CT scans, split into 80/40/40 scans for training, public test, and hidden test cases. VerSe 2020 is a more challenging dataset, containing more CT scans with pathological appearances, fractures, and metal artifacts. We follow the standard data-split setting, with 89/103/103 scans for training, public test, and hidden test cases.

To ensure relatively fixed FOVs of iterative cropped patches in the second stage, we resample all data with a voxel space equal to $1 \times 1 \times 1.998 mm^{3}$ and $0.8262 \times 0.8262 \times 1 mm^{3}$ for VerSe 2019 and 2020 respectively. Then we crop patches containing at least three sequential vertebrae, with the patch size equal to $112 \times 128 \times 64$ and $128 \times 128 \times 96$. Each patch is centered at the spherical centroid of the intermediate vertebra. A random translation is exerted on the patch center to generate the variance in the center regression decoder, with a $[-5, 5]$ range on each coordinate axis. For each middle vertebra, we repeat the translation process 3 times. To further simulate various types of imprecise boundary prediction, different pretrained weights are adopted to generate coarse predictions for the positional prompt. For VerSe 2019, a total of 7767 patches from 80 training CT scans are split as 6214/1553 cases for training and validation. For VerSe 2020, following the same $4\mbox{:}1$ ratio, 10188 patches are split as 8150/2038 cases. During inference, the SLoRD framework is adopted to refine coarse predictions. Moreover, Dice score and Hausdorff distance (HD) are chosen as quantitative metrics \cite{yeghiazaryan2018family}. 

\textbf{Implementation Details.} All experiments are implemented based on Pytorch and trained on 2 NVIDIA Tesla V100 GPUs. In stage 1, MedNeXt \cite{roy2023mednext} is selected as the U-shape backbone with large receptive fields to extract long-term semantics, trained for 1000 epochs with $10.53M$ parameters. And we utilize a combination of cross entropy loss and Dice loss followed by \cite{isensee2021nnu}. The patch size is $128 \times 160 \times 96$. In stage 2, SLoRD is trained for 400 epochs, which employs the encoder of 3D UNet with only $5.15M$ parameters. And we utilize a weighted loss $\mathcal{L}_{total}$ as illustrated in Section \ref{SLoRD}. We train all models using AdamW optimizer. With the linear warm-up strategy, the initial learning rate is set as $5e\mbox{-}4$ with a cosine learning rate decay scheduler, and weight decay is set as $1e\mbox{-}5$. The batch size is 2 and 16 for two stages individually. The hyperparameters $s$ (sampling interval) and $k$ (number of basic contour descriptors) are set as $5^\circ$ and $200$ for a better trade-off between performance and efficiency. 

\begin{table*}[!t]
  \begin{center}
  \caption{Comparisons with other single-stage, two-stage, and multi-stage methods on the public and hidden test datasets of VerSe 2020 (A-Dice: average Dice. M-Dice: median Dice. A-HD: average Hausdorff distance. M-HD: median Hausdorff distance).}
  \label{benchmark verse20}
    \resizebox{0.95\textwidth}{!}{
  \begin{tabular}{clcccccccc}  
  \hline
 \multirow{2}*{Stage} &  \multirow{2}*{Method} & \multicolumn{4}{c}{\textbf{Public test}} & \multicolumn{4}{c}{\textbf{Hidden test}}  \\  
  \cmidrule(r){3-6}    \cmidrule(r){7-10}
  & & A-Dice $\uparrow$ & M-Dice $\uparrow$ & A-HD $\downarrow$ & M-HD $\downarrow$ & A-Dice $\uparrow$ & M-Dice $\uparrow$ & A-HD $\downarrow$ & M-HD $\downarrow$ \\
  \hline  
  & 3D UNet \cite{cciccek20163d} & 82.69  & 89.49  & 11.15  & 9.40  & 83.32 & 88.90 & 10.59 & 9.02  \\
    & Xiangshang Z. \cite{sekuboyina2021verse}  & 83.58 & 92.69 & 15.19 & 9.76 & 85.07 & 93.29 & 12.99 & 8.44 \\
  & nnUNet \cite{isensee2021nnu} & 85.34 & 92.10 & 10.01 & 8.47 & 86.83 & 91.79 & 9.29 & 7.58 \\
  Single-stage &  Verteformer \cite{you2023verteformer} & 85.44 & 92.04 & 9.19 & 8.86 & 86.57 & 91.47  & 9.80 & 8.23 \\ 
  & UNeXt\cite{valanarasu2022unext} & 81.19 & 90.30 & 13.39 & 11.94 & 82.77 & 87.94  & 12.39 & 12.72 \\
  & Swin UNETR \cite{tang2022self} & 82.57 & 90.68  & 12.07 & 10.16 & 81.83 & 88.79  & 12.98 & 11.37 \\
  & MedNeXt \cite{roy2023mednext} & 87.35 & 93.26 & 8.98 & 7.50 & 87.75 & 92.41 & 8.31 & 6.28 \\
  \hline
  Multi-stage & Sekuboyina A. \cite{sekuboyina2021verse} & 78.05 & 85.09 & 10.99 & 6.38 & 79.52 & 85.49 & 11.61 & 7.76 \\ 
  Two-stage & Yeah T. \cite{sekuboyina2021verse} & 88.88 & 92.93 & 9.57 & 5.43 & 87.91 & 92.76 & 8.41 & 5.91 \\ 
  Multi-stage & Payer C. \cite{payer2020coarse} & 91.65  & \underline{95.72} & \underline{5.80} & \underline{4.06} & 89.71 & \textbf{95.65} & \underline{6.06} & \underline{3.94} \\
  Multi-stage & Chen D. \cite{sekuboyina2021verse} & \underline{91.72}  & 95.52 & 6.14 & 4.22 & 91.23 & 95.21 & 7.15 & 4.30 \\
   Two-stage & SAM-Med2D \cite{cheng2023sam}  & 89.03 & 94.31 & 8.76 & 6.51 & 89.74 & 93.48 & 7.95 & 5.28 \\
   Two-stage & Tao R. \cite{tao2022spine}  & 91.29 & 95.33 & 6.30 & 4.73 & \underline{91.65} & 94.72 & 6.29 & 5.07 \\
         \hline 
  Two-stage & Ours   & \textbf{92.09} & \textbf{95.75}  & \textbf{5.30} & \textbf{3.78} & \textbf{92.31}  & \underline{95.33} & \textbf{5.55} & \textbf{3.74} \\
    \hline 
  \end{tabular}}
  \end{center}
\end{table*}
\subsection{Performance Comparison}
\subsubsection{Comparison baselines}
For a comprehensive comparison, we list out eight powerful single-stage baselines, five SOTA two-stage or multi-stage approaches. For single-stage baselines, Lessmann's method \cite{lessmann2019iterative}, Xiangshang's method \cite{sekuboyina2021verse} and Verteformer \cite{you2023verteformer} are specifically devised for the vertebrae segmentation task, while 3D UNet \cite{cciccek20163d}, nnUNet \cite{isensee2021nnu}, UNeXt \cite{valanarasu2022unext}, Swin UNETR \cite{tang2022self}, and MedNeXt \cite{roy2023mednext} serve as general segmentation frameworks feasible for bone-family anatomical structures. Besides, several two-stage and multi-stage methods achieving SOTA performance in the VerSe Challenge are also included in our benchmark, such as Payer's method \cite{payer2020coarse}, Chen's method \cite{sekuboyina2021verse} and Spine-Transformer \cite{tao2022spine}. \changed{M1.3}{\link{R1.3}}{Also, SAM-based methods \cite{cheng2023sam, kirillov2023segment, ma2024segment} show potential in segmenting bone-like anatomical structures. Thus, SAM-Med2D \cite{cheng2023sam} is chosen as a comparative baseline in the benchmark by implementing the axial-plane segmentation. It is worth mentioning that we select a bounding box enclosing vertebrae and a positive point inside vertebrae as prompts. The positive point for each axial slice is set as the centroid of coarsely segmented mask generated from the first stage network. And the label of each segmented vertebra corresponds to the positive point's label in coarse predictions. Thus, SAM-Med2D can be considered a two-stage pipeline.}

\begin{figure}[!t]
\centerline{\includegraphics[width=1.0\linewidth]{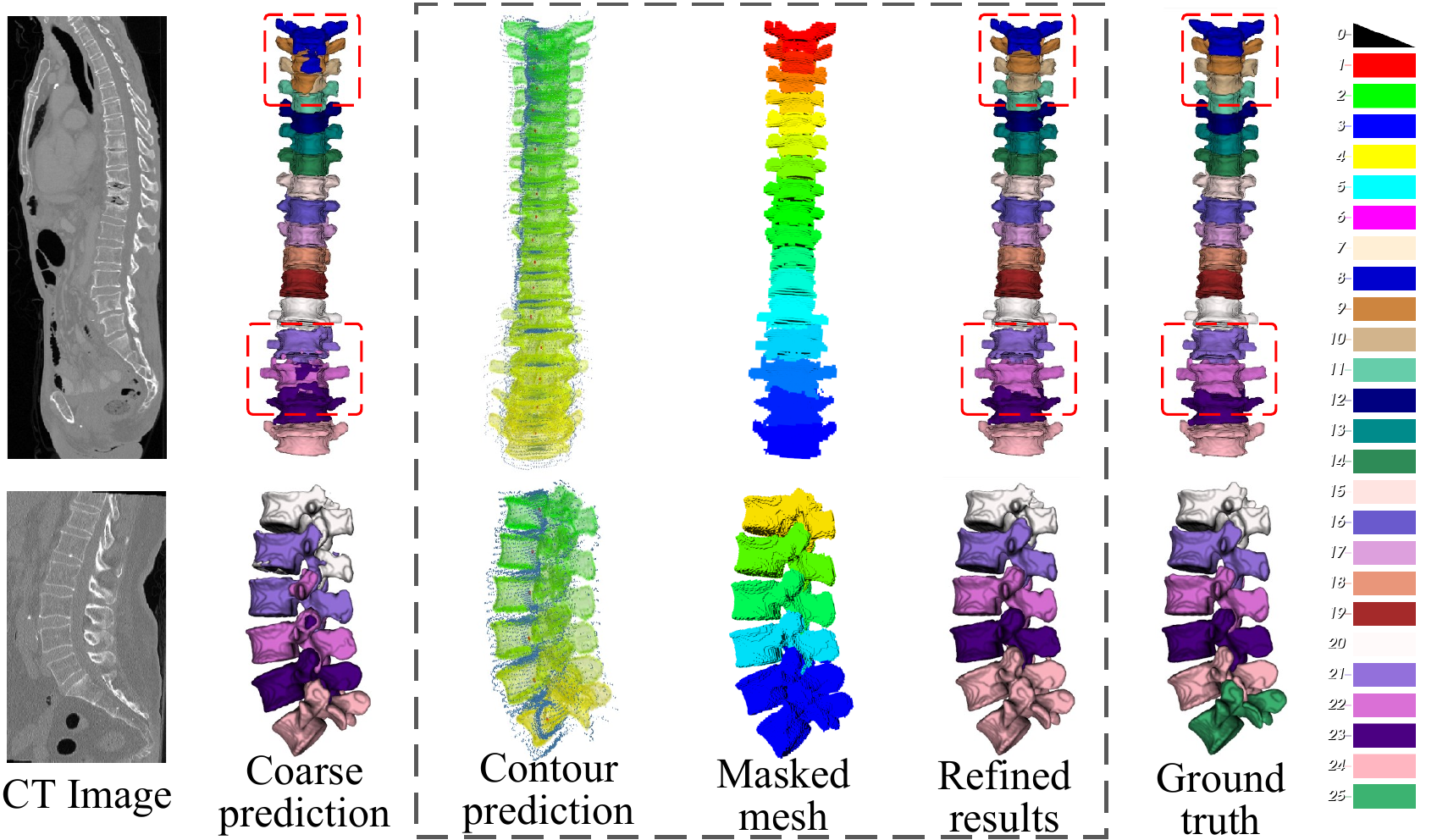}}
\caption{Visualizations of the detailed refinement process. Refined predictions in the dashed box reveal a better segmentation consistency in each vertebra compared with coarse predictions.}
\label{fig4}
\end{figure}

\subsubsection{Results on VerSe 2019}
To evaluate the efficacy of SLoRD, quantitative experimental results are first conducted on VerSe 2019. As illustrated by Table \ref{tab1}, our method achieves SOTA performance on the average Dice and Hausdorff distance (HD) of the hidden test dataset compared with other single-stage, two-stage, and multi-stage methods. Specifically, our framework surpasses MedNeXt \cite{roy2023mednext} (the best single-stage model) by $2.42\%\uparrow$ average Dice score and $5.38mm\downarrow$ average Hausdorff distance. And these two metrics come to $1.39\%\uparrow$ and $0.99mm\downarrow$ when compared with Spine-Transformer \cite{tao2022spine} (the strongest two-stage model), $1.53\%\uparrow$ and $1.11mm\downarrow$ compared with Payer's method \cite{sekuboyina2021verse} (the powerful multi-stage model). \changed{M1.3}{\link{R1.3}}{As shown in Table \ref{tab1}, our model outperforms SAM-Med2D on hidden test datasets of VerSe 2019 with a $2.07\%$ average Dice score increase, a $2.88mm$ average HD decrease.} Moreover, the median Dice and HD of our method also reach quite promising performance.

\subsubsection{Results on VerSe 2020}
To further validate the robustness and universality of SLoRD, we carry out a quantitative analysis on VerSe 2020, which contains more challenging cases with deformed shapes and metal artifacts. As shown in Table \ref{benchmark verse20}, our approach achieves the best boundary delineation results with the lowest average HD values. Compared with Chen's method \cite{sekuboyina2021verse}, $0.84mm\downarrow$ and $1.60mm\downarrow$ Hausdorff distance are acquired for the public and hidden test datasets. Also, the median HD values reveal our method could improve the segmentation performance for not only satisfactory cases but poor cases. Obviously, the superior segmentation performance originates from the strict voxel consistency and contour precision regularization generated from the instance-based SLoRD architecture. By labeling sequential instance masks, we will acquire semantic masks with better consistency. 
\begin{table}[!t]
  \centering
  \caption{The model efficiency evaluation between our method and other approaches on the hidden test datasets of VerSe 2019. The inference speed refers to the average processing time for one CT scan.}
    \resizebox{0.95\columnwidth}{!}{
  \begin{tabular}{lccc}  
  \hline
  Method & Parameters (M) & FLOPs (G) & Inference speed (s)  \\  
  \hline  
    Verteformer \cite{you2023verteformer} & 330.65 & 336.5  & 65.2 \\ 
    nnUNet \cite{isensee2021nnu} & 30.90 & 502.8  & 78.6 \\ 
  Payer C. \cite{payer2020coarse} &  34.33  & 517.6  & 217.9  \\
   Tao R. \cite{tao2022spine}  &  38.41 &  631.3 & 176.3 \\
   Ours  & 25.18 & 454.6 & 142.8 \\
  \hline  
  \end{tabular}}
\label{table 2}
\end{table}

\subsubsection{Qualitative results}
Qualitative visualizations of some challenging cases are displayed in Fig.\ref{benchmark}. The first case visualizes thoracic vertebrae with similar shapes. Besides, the spinous process of the precedent vertebra shows the same height as the superior articular process and vertebral body of the latter vertebra, prone to imprecise boundary predictions. The second case is plagued by metal artifacts, making it difficult to identify and locate several vertebrae in the bottom. The third case illustrates the potential problem of inter-class boundary confusion between normal vertebrae and transitional vertebrae \cite{sekuboyina2021verse}. And our method can better address the above obstacle with intra-vertebrae segmentation consistency.

For a detailed mask refinement process, intermediate results are provided in Fig.\ref{fig4}. Given coarse predictions in stage 1, SLoRD will generate contour predictions for each vertebra. Then discrete contour point clouds are transformed into a sequence of masked meshes. After filling the inner volume, refined masks will be attained with good boundary precision for each class as shown in the dashed box of Fig.\ref{fig4}. SLoRD can function well as long as the top two vertebrae perform satisfactorily, even with moderate inconsistency. It is worth highlighting that each instance mask is generated via the linear combination of basic contour descriptors, which can be acquired by SVD offline. A depiction of decomposed contour priors can be referred to in Fig.\ref{fig2}.

\vspace{0.5mm}
\subsubsection{Model Efficiency}
\changed{M3.7}{\link{R3.7}}{Besides, the efficiency analysis is conducted between our framework and other typical methods on the hidden test dataset of VerSe 2019. Specifically, we choose model parameters, computational costs, and inference time per case as quantitative metrics. As illustrated by Table \ref{table 2}, our approach only contains $25.18$M network parameters, which is the most lightweight model in the referenced methods. Besides, our approach achieves a $142.8$s average inference time for each CT scan, much faster than Payer's method ($217.9$s) and Tao's method ($176.3$s).}

\subsection{Ablation Analysis}
\label{settings for spherical system}\subsubsection{Settings for the spherical-based contour representations} We investigate a detailed ablation analysis on the contour representations based on the spherical coordinate system. For a simple yet effective evaluation, we adopt the average surface distance (ASD) between contour points and the boundary to assess the contour restoration performance. The ASD value in Table \ref{table 3} is calculated via the linear combination of low-rank contour descriptors from SVD. Thus, this distance value stands for the best performance that can be theoretically achieved due to vertebral non-convex shape property and deviation due to integer approximation. 

    \begin{table}[!t]
  \centering
  \caption{Ablation analysis of the rank number, the coordinate system, and the dimension reduction method on the segmentation performance when evaluated on VerSe 2019 and 2020. $k$ means the rank number.}
    \resizebox{0.95\columnwidth}{!}{
  \begin{tabular}{lcccc}
  \hline
 \multirow{2}*{Settings}  & \multicolumn{2}{c}{VerSe 2019} & \multicolumn{2}{c}{VerSe 2020}  \\  
  \cmidrule(r){2-3}    \cmidrule(r){4-5}
   & A-Dice $\uparrow$ & A-HD $\downarrow$ & A-Dice $\uparrow$ & A-HD $\downarrow$ \\  
  \hline  
    Low Rank (k=200) & 91.33 & 5.97  & 92.31 & 5.55 \\ 
   Full Rank (k=2664) &  91.52  & 5.89  & 92.40 & 5.62  \\
  \hline  
    Spherical System & 91.33 & 5.97  & 92.31 & 5.55 \\ 
   Cartesian System &  85.79  & 11.84  & 87.63 & 10.92  \\
  \hline  
    SVD (k=200) & 91.33 & 5.97  & 92.31 & 5.55 \\ 
   PCA (k=200) &  87.61  & 9.49  & 88.27 & 10.60  \\
  \hline 
  \end{tabular}}
\label{r1}
\end{table}

\begin{table}[!t]
  \centering
  \caption{Ablation study on the orientation of positive Z-axis, sampling interval, and the number of basic contour descriptors (ASD \cite{yeghiazaryan2018family}: average surface distance). Bold strings stand for the final settings of SLoRD for the evaluation of segmentation performance.}
    \resizebox{0.90\columnwidth}{!}{
  \begin{tabular}{lccc}  
  \hline 
  \multicolumn{4}{c}{(a) Ablation on the choice of positive Z-axis} \\
  \hline  
  Settings & x-axis & y-axis & \textbf{z-axis}  \\   
  ASD $\downarrow$ & 0.987\scriptsize{$\pm0.074$} & 1.189\scriptsize{$\pm0.114$} & 0.962\scriptsize{$\pm0.073$} \\  
  \hline 
  \multicolumn{4}{c}{(b) Ablation on the sampling interval} \\
    \hline 
    Settings & $3^\circ$ & $\mathbf{5^\circ}$ & $10^\circ$  \\   
    ASD $\downarrow$ & 0.876\scriptsize{$\pm0.067$} & 0.962\scriptsize{$\pm0.073$} & 1.245\scriptsize{$\pm0.122$} \\  
  \hline 
    \multicolumn{4}{c}{(c) Ablation on the rank number} \\
  \hline 
    Settings & 100 & \textbf{200} & 500  \\   
    ASD $\downarrow$ & 1.033\scriptsize{$\pm0.080$} & 0.962\scriptsize{$\pm0.073$} & 0.917\scriptsize{$\pm0.068$} \\  
  \hline 
  \end{tabular}}
\label{table 3}
\end{table}

\begin{table*}[!t]
  \begin{center}
  \caption{Ablation study on the combination of decoders on VerSe 2020. $D_{coe}$: coefficient decoder, $D_{cen}$: center decoder, $D_{mask}$: mask decoder, Prompts: positional Gaussian prompts.}
  \label{ablation_decoders}
  \resizebox{1.0\textwidth}{!}{
  \begin{tabular}{cccccccccccc}
  \hline  
  \multirow{2}*{$D_{coe}$} &  \multirow{2}*{$D_{cen}$} &  \multirow{2}*{$D_{mask}$} &  \multirow{2}*{Prompts} & \multicolumn{4}{c}{Dice score (\%) $\uparrow$} & \multicolumn{4}{c}{$HD$ (mm) $\downarrow$}   \\  
  \cmidrule(r){5-8}  \cmidrule(r){9-12}
  & & & & Cervical & Thoracic & Lumbar & Mean & Cervical & Thoracic & Lumbar & Mean  \\
  \hline  
  \textcolor{red}{\ding{55}} & \textcolor{green}{\ding{52}} & \textcolor{red}{\ding{55}} & \textcolor{red}{\ding{55}} & - & - & - & - & - & - & - & - \\
  \textcolor{green}{\ding{52}} & \textcolor{red}{\ding{55}} & \textcolor{red}{\ding{55}} & \textcolor{red}{\ding{55}} & 80.09 & 82.46 & 79.65 & 81.39 & 19.37 & 16.59 & 17.83 & 17.54 \\
  \textcolor{green}{\ding{52}} & \textcolor{green}{\ding{52}} & \textcolor{red}{\ding{55}} & \textcolor{red}{\ding{55}} & 87.90 & 88.75 & 86.68 & 87.86 & 10.39 & 8.60 & 9.43 & 9.03 \\ 
  \textcolor{green}{\ding{52}} & \textcolor{green}{\ding{52}} & \textcolor{red}{\ding{55}} & \textcolor{green}{\ding{52}} & 87.82 & 90.32 & 88.38 & 90.13 & 8.00 & 6.87 & 7.07 & 7.29   \\
    \textcolor{green}{\ding{52}} & \textcolor{green}{\ding{52}} & \textcolor{green}{\ding{52}} & \textcolor{red}{\ding{55}} & 89.58 & 89.83 & 89.62 & 91.22 & 6.86 & 7.09 & 6.19 & 6.63 \\
  \textcolor{green}{\ding{52}} & \textcolor{red}{\ding{55}} & \textcolor{green}{\ding{52}} & \textcolor{green}{\ding{52}} & 87.41 & 90.50 & 85.96 & 88.64 & 8.13 & 6.14 & 9.11 & 8.09   \\
  \textcolor{green}{\ding{52}} & \textcolor{green}{\ding{52}} & \textcolor{green}{\ding{52}} & \textcolor{green}{\ding{52}} &  \textbf{90.27} &  \textbf{93.44} &  \textbf{90.69} &  \textbf{92.31} &   \textbf{6.37} &  \textbf{4.68} &  \textbf{6.04} &  \textbf{5.55}  \\
  \hline  
  \end{tabular}}
  \end{center}
\end{table*}

\begin{figure}[!t]
\centerline{\includegraphics[width=0.95\linewidth]{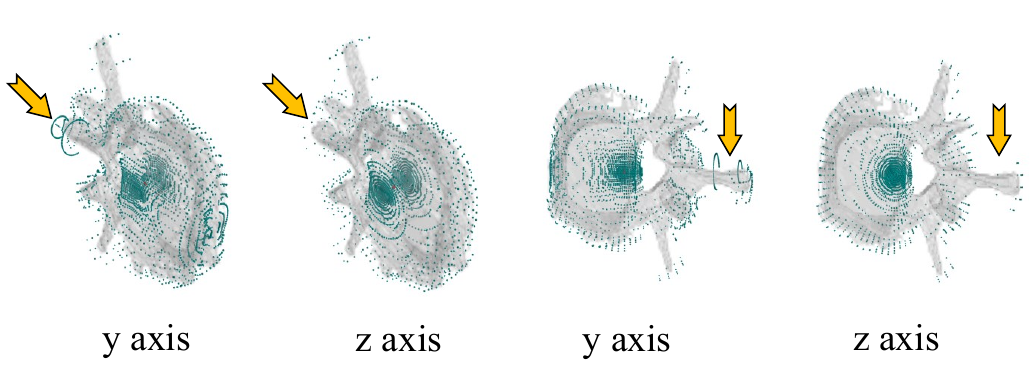}}
\caption{Visualizations of the full-rank contour restoration on the orientation setting of positive Z-axis in the spherical coordinate system. The voxel z axis from inferior to superior orientations is chosen as the final experimental setting.}
\label{contour_axis}
\end{figure}

\textbf{The orientation of the positive Z-axis in the spherical coordinate system:} As revealed in Table \ref{table 3}, the voxel z-axis as the objective orientation achieves the lowest ASD value. Compared with the voxel x-axis, there is a $0.227$ voxel distance drop. Besides, as revealed in Fig.\ref{contour_axis}, there will be a deviation for contour points adjacent to the spinous process of vertebrae when comparing the setting of z-axis with y-axis. 

\textbf{Sampling interval of the azimuth angle $\phi$ and the polar angle $\theta$:} Different values of the angular interval are also considered. As shown in Table \ref{table 3}, a smaller interval means a denser sampling for the contour point cloud, which will promote the precision of regressed contours by SLoRD, and then elevate the final segmentation performance. However, that behavior will inevitably bring heavier computational costs. Considering the balance, that hyper-parameter is specified as $5^{\circ}$, then the dimension $N$ of each contour descriptor $\bm{\rho}$ is equal to $\tfrac{360^{\circ}}{5^{\circ}} \times (\tfrac{180^{\circ}}{5^{\circ}} + 1) = 2664$.

\textbf{The rank of basic contour descriptors:} According to Table \ref{table 3}, $200$ basic contour descriptors can already restore the original contour with sufficient precision. That is also proven in the qualitative results of Fig.\ref{fig2}. Considering both efficacy and efficiency, the rank number is set as $200$ instead of $500$. Moreover, we report the segmentation performance of our model with the guidance of low-rank ($k=200$) and full-rank ($k=2664$) contour descriptors. As shown in Table \ref{r1}, by replacing the setting of full-rank with low-rank, the average Dice score only decreases $0.09\%$ for VerSe 2020.

\textbf{Spherical centroid vs. Centroid:} Adopting the spherical centroid reduces the ASD value in both low-rank and full-rank approximations, with a $0.053\downarrow$ and $0.267\downarrow$ respectively as shown in Fig.\ref{fig-comparison}. That demonstrates the advancement of spherical centroids owing to the distance-based optimization objective between contour points and the vertebral surface.

\changed{M3.6}{\link{R3.6}}{\textbf{Spherical coordinate system vs. Cartesian coordinate system:} To prove the efficacy of the spherical system in vertebral contour representations, we further conduct a comparison study on the type of coordinate systems. For a fair comparison, the number of sampled contour points for shape completion is unified as $2664$. As indicated by Table \ref{r1}, the segmentation performance of SLoRD conditioned on the Cartesian coordinate system is significantly inferior to that conditioned on the spherical system. As the methodology part states, the spherical coordinate system introduces implicit angular priors corresponding to $2664$ directions. Thus, networks can better reconstruct shape representations via basic contour descriptors.}

\changed{M4.4.2}{\link{R4.4}}{\textbf{PCA vs. SVD:} Since spherical-based basic descriptors are utilized to boost the final segmentation performance, a quantitative analysis on the way of extracting basic contour descriptors is carried out. Specifically, our adopted SVD strategy reveals a stronger numerical stability and accuracy than PCA. As shown in Table \ref{r1}, there are $3.72\%\uparrow$ and $4.04\%\uparrow$ average Dice score for hidden test datasets of VerSe 2019 and 2020. That phenomenon demonstrates that SVD is better represent nonlinear high-dimensional vertebral data.}

\begin{figure}[!t]
\centerline{\includegraphics[width=1.0\linewidth]{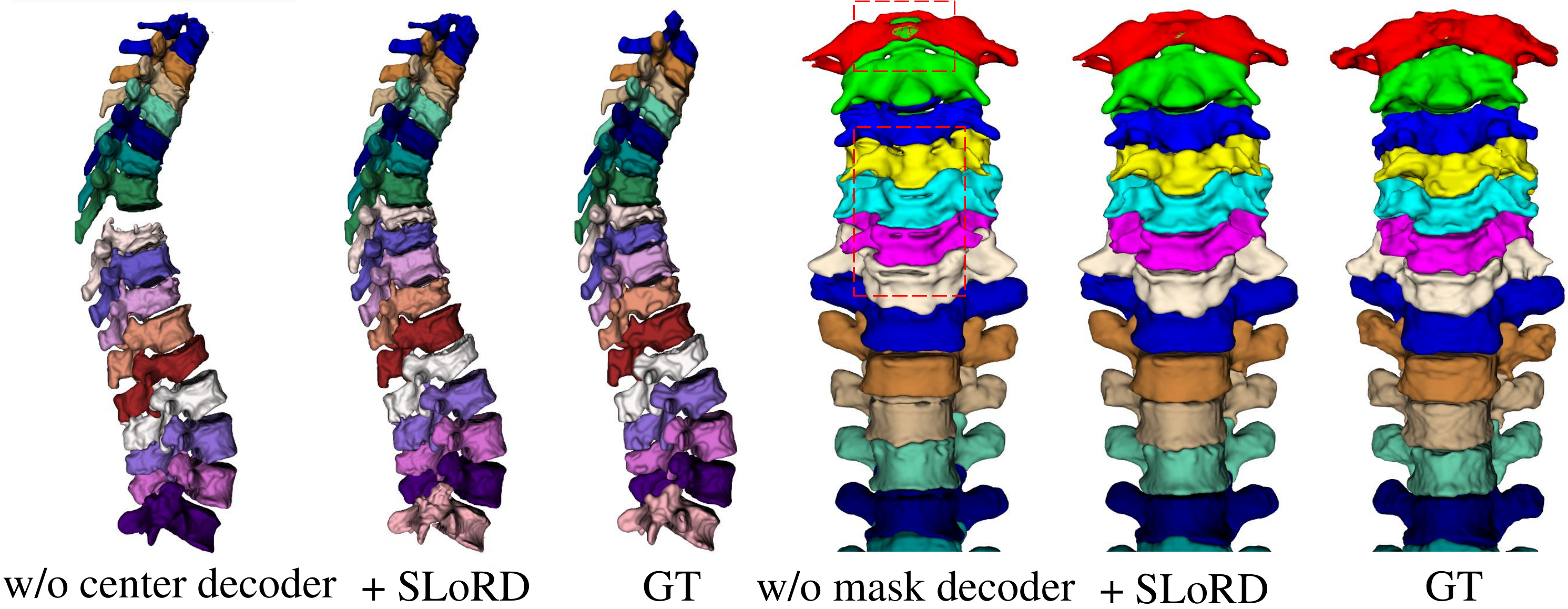}}
\caption{Qualitative results for the ablation study on the center decoder and mask decoder of SLoRD.}
\label{structural_ablation}
\end{figure}

\subsubsection{Structural ablation on SLoRD} 
Furthermore, we conduct the structural ablation analysis for SLoRD on VerSe 2020. 

\textbf{Coefficient decoder:} Without the coefficient decoder, contours cannot be generated. Thus, that is the essential part. For the reason why the regression of contour points is not performed directly in this part, we reckon that models show deficient abilities to perceive and localize contour points with the specific angle. Therefore, basic contour descriptors with latent angular information serve as strong guidance to generate target contour points.

    \begin{table}[!t]
  \centering
  \caption{Analysis on the plug-and-play property of SLoRD when applied to other methods on vertebrae segmentation. Quantitative results on the hidden test dataset of VerSe 2019 and 2020.}
    \resizebox{0.95\columnwidth}{!}{
  \begin{tabular}{lcccc}  
  \hline  
  \multicolumn{5}{c}{VerSe 2019} \\
    \hline  
  Methods & A-Dice $\uparrow$ & M-Dice $\uparrow$  & A-HD $\downarrow$ & M-HD $\downarrow$ \\  
  \hline  
  Verteformer \cite{you2023verteformer} & 86.54 & 90.74 & 10.55 & 10.51 \\
  Verteformer \cite{you2023verteformer} + SLoRD & \textbf{89.34} & \textbf{92.98} & \textbf{6.73} & \textbf{6.31} \\
  \hline
  3D UNet \cite{cciccek20163d} & 81.28 & 87.54 & 16.35 & 14.73 \\
  3D UNet \cite{cciccek20163d} + SLoRD & \textbf{85.11} & \textbf{87.49} & \textbf{14.76} & \textbf{11.08} \\
    \hline
  nnUNet \cite{isensee2021nnu} & 86.59 & 92.62 & 12.78 & 12.00 \\
  nnUNet \cite{isensee2021nnu} + SLoRD & \textbf{88.60} & \textbf{90.37} & \textbf{9.05} & \textbf{8.52} \\
  \hline  
  Payer C. \cite{payer2020coarse} & 89.80 & \textbf{95.47} & 7.08 & 4.45 \\
  Payer C. \cite{payer2020coarse} + SLoRD & \textbf{90.97} & 95.26 & \textbf{5.93} & \textbf{4.28} \\
  \hline
  Swin UNETR \cite{tang2022self} & 83.46 & 88.91 & 16.03 & 13.58 \\
   Swin UNETR \cite{tang2022self} + SLoRD & \textbf{87.43} & \textbf{89.24} & \textbf{10.76} & \textbf{8.74} \\
    \hline  
  Tao R. \cite{tao2022spine} & 89.94 & 94.01 & 6.96 & 4.73 \\
  Tao R. \cite{tao2022spine} + SLoRD & \textbf{91.46} & \textbf{95.72} & \textbf{5.73} & \textbf{4.23} \\
  \hline  
    \multicolumn{5}{c}{VerSe 2020} \\
    \hline  
      Verteformer \cite{you2023verteformer} & 86.57 & 91.47 & 9.80 & 8.23 \\
  Verteformer \cite{you2023verteformer} + SLoRD & \textbf{89.22} & \textbf{93.90} & \textbf{7.64} & \textbf{5.55} \\
  \hline  
  3D UNet \cite{cciccek20163d} & 83.32 & 88.90 & 10.59 & 9.02 \\
  3D UNet \cite{cciccek20163d} + SLoRD & \textbf{86.16} & \textbf{87.97} & \textbf{10.39} & \textbf{9.85} \\
    \hline
  nnUNet \cite{isensee2021nnu} & 86.83 & 91.79 & 9.29 & 7.58 \\
  nnUNet \cite{isensee2021nnu} + SLoRD & \textbf{88.07} & \textbf{91.95} & \textbf{8.47} & \textbf{7.11} \\
  \hline  
  Payer C. \cite{payer2020coarse} & 89.71 & 95.65 & 6.06 & \textbf{3.94} \\
  Payer C. \cite{payer2020coarse} + SLoRD & \textbf{91.72} & \textbf{95.73} & \textbf{5.58} & 4.08 \\
    \hline  
      Swin UNETR \cite{tang2022self} & 81.83 & 88.79 & 12.98 & 11.37 \\
   Swin UNETR \cite{tang2022self} + SLoRD & \textbf{86.44} & \textbf{90.52} & \textbf{9.35} & \textbf{8.30} \\
    \hline  
  Tao R. \cite{tao2022spine} & 91.65 & 94.72 & 6.29 & 5.07 \\
  Tao R. \cite{tao2022spine} + SLoRD & \textbf{92.24} & \textbf{95.11} & \textbf{5.27} & \textbf{4.34} \\
  \hline  
  \end{tabular}}
\label{plug-and-play}
\end{table}

\textbf{Center decoder:} Since the coupling effects between spherical centers and coefficients, adding the center decoder to the coefficient decoder will enhance the segmentation performance significantly, with $6.47\%\uparrow$ average Dice score and $8.51mm\uparrow$ average HD as shown in Table \ref{ablation_decoders}. Besides, as portrayed in Fig.\ref{structural_ablation}, there exists a missing detection for a specific vertebra, resulting in sequential identification errors. That phenomenon significantly weakens segmentation results on lumbar vertebrae \cite{chen2019vertebrae}. 

\textbf{Mask decoder:} Without it, SLoRD reveals degraded segmentation performance, with a $2.18\%\downarrow$ average Dice score according to Table \ref{ablation_decoders}. Carrying out the cause and effect analysis, this mask decoder will bring additional shape information, which boosts the contour restoration precision according to Eq \eqref{rank-k}. For more detailed evaluations, the missing of this component will affect the boundary delineation for cervical vertebrae as shown in Fig.\ref{structural_ablation}.

\textbf{Positional prompts:} The positional priors promote the regression for spherical centroids, further boost the precision of contour generation. This detection guidance is beneficial to segment the thoracic region with adjacent vertebrae. As revealed in Table \ref{ablation_decoders}, by adding positional prompts to three decoders, the segmentation performance is boosted with $1.09\%\uparrow$ Dice score and $1.08mm\downarrow$ HD.

\begin{figure}[!t]
\centerline{\includegraphics[width=1.0\linewidth]{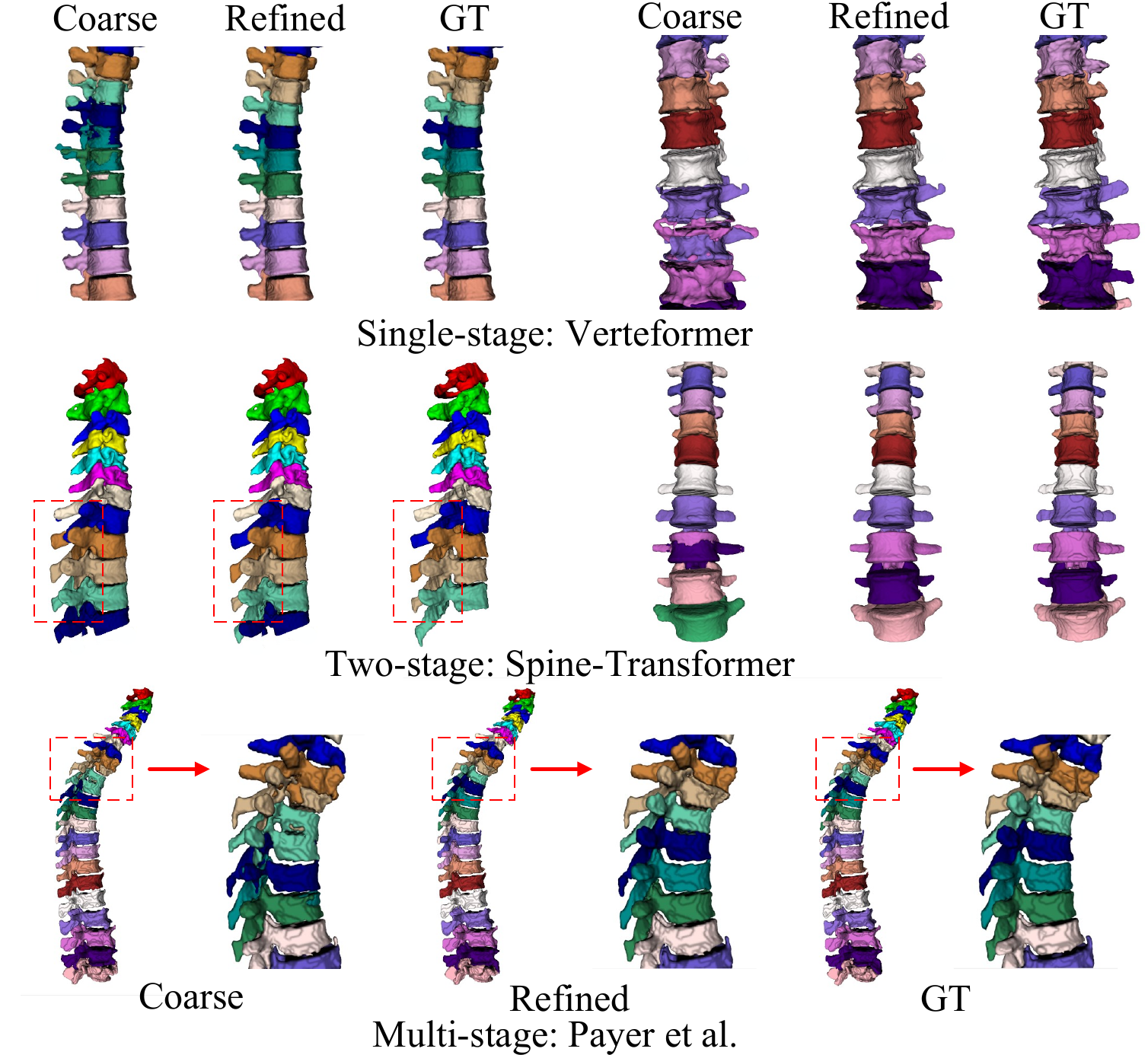}}
\caption{Improvement of our proposed SLoRD when applied to coarse predictions from other segmentation networks including the single-stage Verteformer \cite{you2023verteformer}, two-stage Spine-Transformer \cite{tao2022spine}, and multi-stage Payer's method \cite{payer2020coarse}.}
\label{slord_refine}
\end{figure}

\subsubsection{Analysis on the plug-and-play property} \changed{M3.4}{\link{R3.4}}{We also carried out the ablation study on the plug-and-play property of SLoRD on the hidden test dataset of VerSe 2019 and 2020. Besides MedNeXt \cite{roy2023mednext}, the proposed SLoRD will boost segmentation results achieved by 3D UNet \cite{cciccek20163d}, nnUNet \cite{isensee2021nnu}, Verteformer \cite{you2023verteformer}, Swin UNETR \cite{tang2022self}, Spine-Transformer \cite{tao2022spine}, Payer's method \cite{payer2020coarse} as shown in Table \ref{plug-and-play}. Specifically, for Verteformer on VerSe 2020, substantial improvements have been achieved on the average Dice score and HD ($2.65\%\uparrow$, $2.16mm\downarrow$). Also, SLoRD can repair segmentation results by Payer's approach, with $2.01\%\uparrow$ average Dice score and $0.48mm\downarrow$ average HD. These two metrics reach $0.59\%\uparrow$ and $1.02mm\downarrow$ for Tao's method.} Qualitative results are provided in Fig.\ref{slord_refine} to demonstrate the efficacy of SLoRD in promoting segmentation consistency when plugged into single-stage, two-stage, and multi-stage methods. 

\begin{figure}[!t]
  \centering
 
\begin{minipage}{.48\linewidth}
 \centerline{\includegraphics[width=1.0\linewidth]{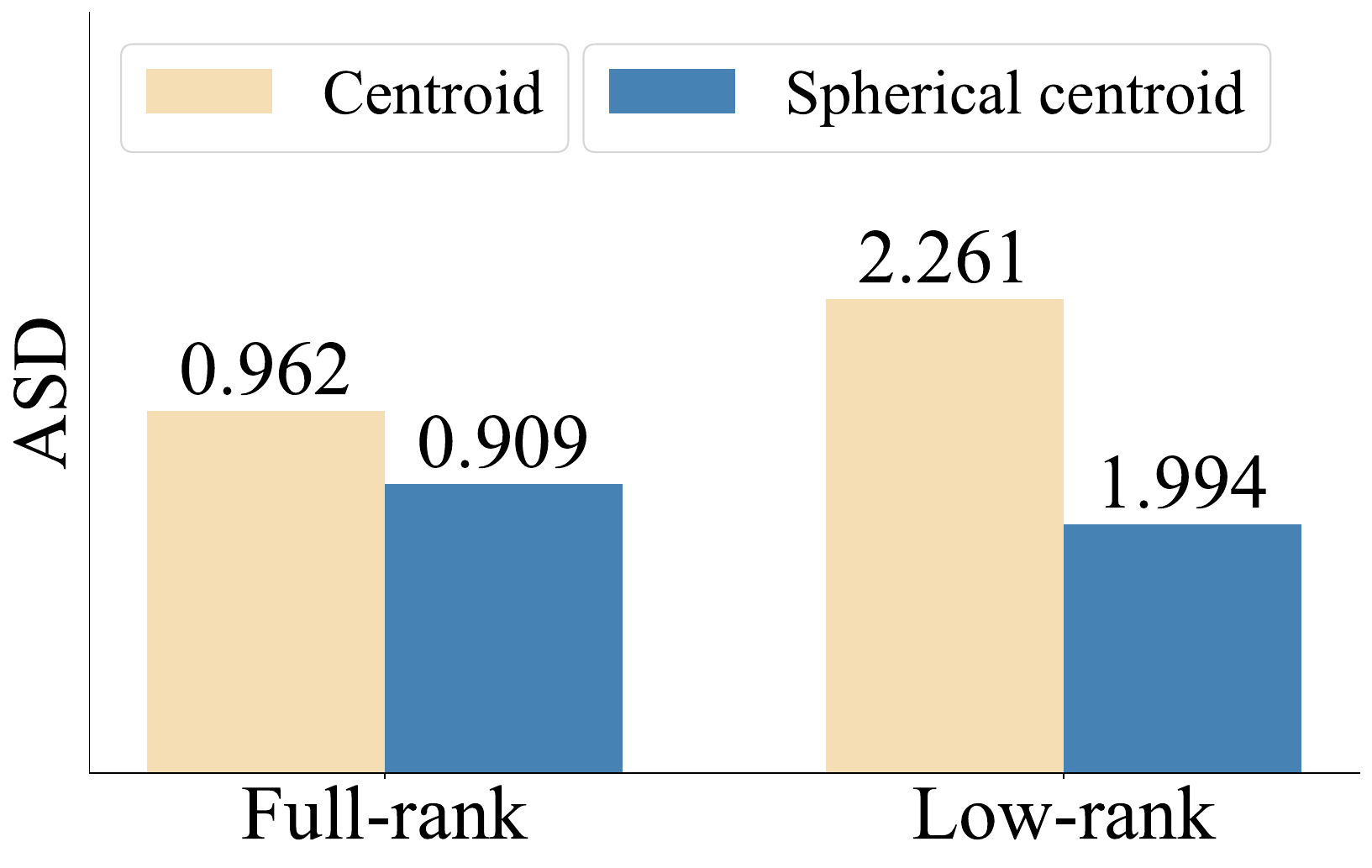}}
\end{minipage}
\hfill
\begin{minipage}{0.48\linewidth}
 \centerline{\includegraphics[width=1.0\linewidth]{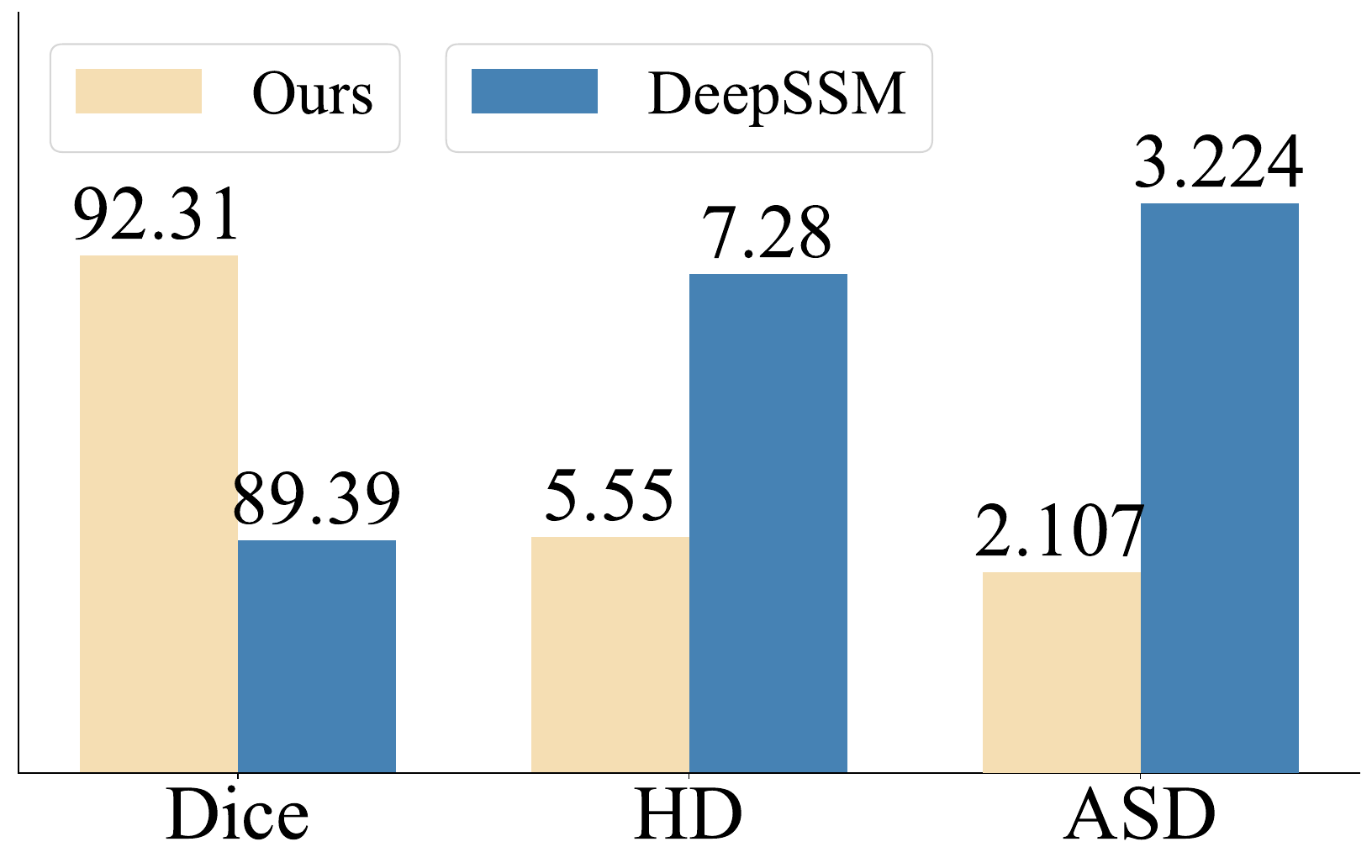}}
\end{minipage}
\caption{Left: ablation study on the type of centers between the centroid and spherical center (ASD \cite{yeghiazaryan2018family}: average surface distance).  Right: performance comparison between our method and DeepSSM on VerSe 2020.}
\label{fig-comparison}
\end{figure}

\begin{figure*}[!t]
\centerline{\includegraphics[width=1.0\linewidth]{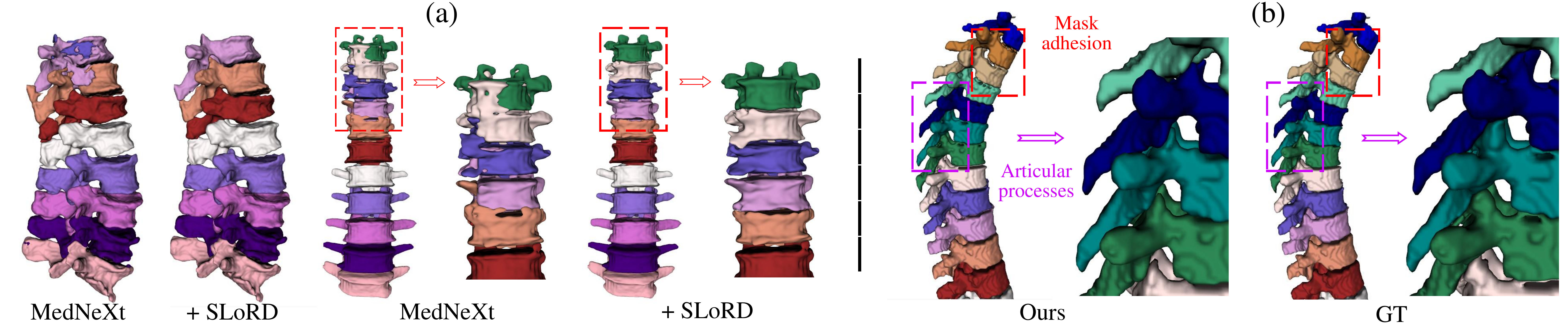}}
\caption{(a) The mask refinement results by SLoRD even with segmentation inconsistency in the topmost vertebra. (b) The limitations of SLoRD. Red: mask adhesion between thoracic vertebrae. Yellow: Imprecise contour regression for the articular processes.}
\label{robustness_limitation}
\end{figure*}

\subsection{Discussion}
\subsubsection{The challenge of similar appearances, various pathologies and metal artifacts}
Talking about how to address the identification and segmentation of vertebrae with similar appearances, our proposed SLoRD adopts sequential modeling for three consecutive vertebrae. The introduction of Gaussian positional prompts helps to distinguish vertebrae with similar appearances. Besides, basic contour descriptors further boost the precision of contour regression. By labeling sequential binary masks, we could realize precise vertebrae segmentation with arbitrary FOVs.

For pathological vertebrae with deformed shapes, coarse predictions in the first stage may not ensure accurate geometric centers. However, as mentioned in the details of the center decoder, coarse centers will be refined as spherical centroids under the regularization of $\mathcal{L}_{center}$ as revealed in Fig.\ref{fig3}. Additionally, the coupling between spherical centroids and linear coefficients is strongly correlated with the vertebral shape. By jointly optimizing spherical centers and contour points, SLoRD is capable to better detect geometric centers of vertebrae even with various pathologies.

\changed{M1.5}{\link{R1.5}}{Besides, for the precise segmentation of CT scans with metal artifacts, existing methods may result in segmentation inconsistency on the exact vertebrae with metal artifacts as depicted by Fig.\ref{benchmark} and Fig.\ref{fig1}. But the positive side is that they can depict precise binary masks, not covering regions with metal artifacts. Here according to the mechanism of devised contour representations, the spherical system chooses the maximum distance at the same angle. Thus, when there exist holes inside the specific vertebra, the contour descriptor is robust enough to represent the exterior contour information, despite metal artifact-corrupted regions also being included. Thus, by combining coarse predictions by the first-stage network and labeled results by SLoRD, we will acquire refined segmentation results with both shape precision and excellent intra-vertebrae consistency.}

\subsubsection{Comparison with iterative fully convolutional network}
We have mentioned that our proposed pipeline is superior to existing methods due to the synchronous introduction of contour precision and explicit consistency constraints. Herein we focus on the analysis of Lessmann's single-stage pipeline \cite{lessmann2019iterative} which is based on the similar strategy of iteratively sliding 3D patches for parallel tasks of vertebral detection and segmentation. According to the analysis from recent works \cite{tao2022spine} and \cite{meng2023vertebrae}, Lessmann's method heavily relies on the first detection, and the failure to find the first segment may result in the failure to precisely detect all subsequent vertebrae. In fact, the iterative framework \cite{lessmann2019iterative} determines the location of the topmost vertebrae using a sliding window fashion with a constant step size. Hence, the choice for the step size is tricky to acquire the first patch precisely centered at the topmost vertebra.

    \begin{table}[!t]
  \centering
  \caption{Comparisons with other single-stage, two-stage, and multi-stage methods on the SPIDER MR dataset. It is worth to mention that our work is aimed at the segmentation task for bone-like structure. Thus, only the shape of vertebrae anatomies are used for evaluation, intervertebral discs and spinal cord not included.}
    \resizebox{1.0\columnwidth}{!}{
  \begin{tabular}{lcccc}  
  \hline
 \multirow{2}*{Method}  & \multicolumn{4}{c}{SPIDER MRI}  \\  
  \cmidrule(r){2-5}
       & A-Dice $\uparrow$ & A-HD $\downarrow$ & M-Dice $\uparrow$ & M-HD $\downarrow$ \\  
  \hline  
    3D UNet \cite{cciccek20163d}  & 91.96 & 6.89 & 94.40 & 5.25  \\ 
   nnUNet \cite{isensee2021nnu} &  92.84  & 4.97  & 95.27 & 4.38  \\
   Verteformer \cite{you2023verteformer} &  92.60  & 5.93  & 95.03 & 4.83  \\
    UNeXt \cite{valanarasu2022unext}  & 92.51 & 6.27 & 94.93 & 5.46  \\ 
   Swin UNETR \cite{tang2022self} &  92.44  & 6.52  & 94.61 & 5.21  \\
   MedNeXt \cite{roy2023mednext} &  \underline{93.48}  & 4.55  & 95.76 & 3.89  \\
    Payer C. \cite{payer2020coarse}  &  93.19  & 4.28 & 95.60 & 3.74  \\ 
   Tao R. \cite{tao2022spine} &  93.46  & \underline{3.95} & \underline{95.87} & \underline{3.36}  \\
   Ours &   \textbf{94.75}  &  \textbf{3.11}  &  \textbf{96.39} &  \textbf{2.58}  \\   
  \hline  
  \end{tabular}}
\label{table spider}
\end{table}
In comparison, our two-stage framework detects the topmost vertebra via the label distribution from predictions in the first stage, which can provide sufficient positional information. In the second stage, SLoRD can jointly optimize the vertebral detection and segmentation tasks. The interaction between the spherical center and contour points will refine the detection precision of vertebrae and boost the segmentation performance in the meantime. Thus, even if there is a label inconsistency in the top two vertebrae as shown in Fig.\ref{robustness_limitation}(a), SLoRD is robust enough to output segmentation masks with good consistency.

\subsubsection{Comparison with PCA-based shape model} 
As we mentioned in Section \ref{contour descriptors}, SVD shows the best low-rank approximation property for high-dimensional nonlinear data and can mine more robust shape descriptors compared with PCA. Here a straight comparison is implemented on the hidden test dataset of VerSe 2020. As illustrated in Fig.\ref{fig-comparison}, our method outperforms DeepSSM significantly, with a $3.92\%$ higher Dice score and a $1.73mm$ lower Hausdorff distance. Also, since different vertebrae bear various numbers of contour points, it is tricky to carry out the uniform sampling process while delineating the original contour. In contrast, our method achieves a $1.117\downarrow$ ASD value, showing more satisfactory contour reconstruction results based on the structural encoding way of contour representations.

\subsubsection{Generalizability on datasets with diverse modalities} 
\changed{M3.8}{\link{R3.8}}{To further demonstrate the generalization ability of SLoRD, we conduct experiments on diverse modalities of vertebral data. SPIDER is an MRI vertebral dataset different from VerSe datasets with the CT modality. As shown in Table \ref{table spider}, our proposed two-stage model outperforms Spine-Transformer \cite{tao2022spine} by $1.29\%\uparrow$ average Dice score and $0.84mm\downarrow$ average Hausdorff distance. Additionally, our model is superior to Payer's method with $1.56\%\uparrow$ average Dice score and $1.17mm\downarrow$ average Hausdorff distance. Also, the median Dice and HD values of our model reach the best among the benchmark. Therefore, SLoRD reveals quite a promising generalization ability on vertebral data with different modalities.}

\subsubsection{The number of vertebrae to address in a sliding window by SLoRD} 
\changed{M4.8}{\link{R4.8}}{Since a sequential label information naturally exists inside the spine, three vertebrae are involved into the refinement process by SLoRD to boost the identification of vertebrae. Therefore, we conduct an ablation study on the number of vertebrae processed in a sliding window. As shown in Table \ref{ablation-vertebrae number}, when three vertebrae are addressed in each sliding window, we can achieve the best segmentation performance, with $2.93\%\uparrow$ average Dice score compared with the setting of 1-vertebra. Besides, the setting of 2-vertebrae achieves comparable segmentation performance compared to that of 3-vertebrae. Indeed, the setting of 2-vertebrae processed in a sliding window also works due to the existence of sequential label priors.}

    \begin{table}[!t]
  \centering
  \caption{Performance analysis on the number of vertebrae processed each time in a sliding window by SLoRD when evaluated on 2020.}
    \resizebox{0.95\columnwidth}{!}{
  \begin{tabular}{lcccc}  
  \hline
 \multirow{2}*{Settings}  & \multicolumn{4}{c}{VerSe 2020}  \\  
  \cmidrule(r){2-3}    \cmidrule(r){4-5}
       & A-Dice $\uparrow$ & A-HD $\downarrow$ & M-Dice $\uparrow$ & M-HD $\downarrow$ \\  
  \hline  
    3-vertebrae  & \textbf{92.31} & \textbf{5.55} & \textbf{95.33} & \textbf{3.74}  \\ 
   2-vertebrae &  92.10  & 5.95  & 95.17 & 3.90  \\
   1-vertebra &  89.38  & 7.73  & 94.96 & 4.26  \\
  \hline  
  \end{tabular}}
\label{ablation-vertebrae number}
\end{table}
\subsubsection{Limitations}
\label{limitation}
Although the proposed SLoRD could achieve segmentation results with better intra-vertebrae consistency and contour precision. There still exists segmentation errors for the superior and inferior articular processes, especially in the region of with large curvature changes. Qualitative results are visualized in Fig.\ref{robustness_limitation}(b). Moreover, there also might be a slight mask adhesion in the vertebral body, especially the thoracic vertebrae which show a close distance to each other. Under that circumstance, a uniform sampling degree cannot meet the demand for the shape modeling with complex shapes. Instead, we reckon that a curvature-based sampling strategy should be considered to deal with abrupt boundary changes in the future work.

\section{Conclusion}
\changed{M5.3}{\link{R5.3}}{In this work, to mitigate segmentation inconsistencies within vertebrae caused by the low contrast between neighboring vertebral structures and diverse pathological manifestations, we introduce SLoRD (Spherical-coordinate-based Local Regularization and Detection), a novel framework that simultaneously enforces voxel-level uniformity and boundary refinement constraints. Specifically, we develop an innovative structural contour encoding method grounded in spherical coordinate transformation. Additionally, fundamental contour descriptors coupled with anatomical shape priors guide the regressed boundary points to closely adhere to the vertebral surfaces. Comprehensive quantitative and qualitative evaluations on the VerSe 2019 and 2020 datasets confirm the superior performance of our approach over existing single-stage, two-stage, and multi-stage segmentation models. Furthermore, SLoRD functions as a versatile plug-and-play module, effectively correcting segmentation inconsistencies in coarse predictions generated by other methods. As for the limitation of SLoRD on segmenting regions with large curvatures, the future research is required by potentially adopting the curvature-based sampling strategy, which will further boost the segmentation performance.}

\section*{Acknowledgment}
Conflict of Interest: The authors declare that they have no
conflict of interest.

\section*{References}
\vspace{-2mm}
\bibliographystyle{abbrv}  
\bibliography{main}

\begin{thebibliography}{10}

\bibitem{lessmann2019iterative}
N.~Lessmann {\em et~al.}, ``Iterative fully convolutional neural networks for automatic vertebra segmentation and identification,'' {\em Medical image analysis}, vol.~53, pp.~142--155, 2019.

\bibitem{burns2016automated}
J.~E. Burns, J.~Yao, {\em et~al.}, ``Automated detection, localization, and classification of traumatic vertebral body fractures in the thoracic and lumbar spine at ct,'' {\em Radiology}, vol.~278, no.~1, pp.~64--73, 2016.

\bibitem{knez2016computer}
D.~Knez, B.~Likar, F.~Pernu{\v{s}}, and T.~Vrtovec, ``Computer-assisted screw size and insertion trajectory planning for pedicle screw placement surgery,'' {\em IEEE TMI}, vol.~35, no.~6, pp.~1420--1430, 2016.

\bibitem{frost2019materials}
B.~A. Frost, S.~Camarero-Espinosa, and E.~J. Foster, ``Materials for the spine: anatomy, problems, and solutions,'' {\em Materials}, vol.~12, no.~2, p.~253, 2019.

\bibitem{naegel2007using}
B.~Naegel, ``Using mathematical morphology for the anatomical labeling of vertebrae from 3d ct-scan images,'' {\em Computerized Medical Imaging and Graphics}, vol.~31, no.~3, pp.~141--156, 2007.

\bibitem{sekuboyina2021verse}
A.~Sekuboyina, M.~E. Husseini, {\em et~al.}, ``Verse: a vertebrae labelling and segmentation benchmark for multi-detector ct images,'' {\em Medical image analysis}, vol.~73, p.~102166, 2021.

\bibitem{tao2022spine}
R.~Tao, W.~Liu, and G.~Zheng, ``Spine-transformers: Vertebra labeling and segmentation in arbitrary field-of-view spine cts via 3d transformers,'' {\em Medical Image Analysis}, vol.~75, p.~102258, 2022.

\bibitem{mao2024semantics}
Y.~Mao, Q.~Feng, Y.~Zhang, and Z.~Ning, ``Semantics and instance interactive learning for labeling and segmentation of vertebrae in ct images,'' {\em Medical Image Analysis}, p.~103380, 2024.

\bibitem{you2024learning}
X.~You, J.~He, J.~Yang, and Y.~Gu, ``Learning with explicit shape priors for medical image segmentation,'' {\em IEEE TMI}, 2024.

\bibitem{you2023verteformer}
X.~You, Y.~Gu, Y.~Liu, S.~Lu, X.~Tang, and J.~Yang, ``Verteformer: A single-staged transformer network for vertebrae segmentation from ct images with arbitrary field of views,'' {\em Medical Physics}, vol.~50, no.~10, pp.~6296--6318, 2023.

\bibitem{klinder2009automated}
T.~Klinder, J.~Ostermann, M.~Ehm, A.~Franz, R.~Kneser, and C.~Lorenz, ``Automated model-based vertebra detection, identification, and segmentation in ct images,'' {\em MedIA}, vol.~13, no.~3, pp.~471--482, 2009.

\bibitem{hammernik2015vertebrae}
K.~Hammernik, T.~Ebner, D.~Stern, M.~Urschler, and T.~Pock, ``Vertebrae segmentation in 3d ct images based on a variational framework,'' {\em Recent advances in computational methods and clinical applications for spine imaging}, pp.~227--233, 2015.

\bibitem{lim2014robust}
P.~H. Lim, U.~Bagci, and L.~Bai, ``A robust segmentation framework for spine trauma diagnosis,'' in {\em Computational Methods and Clinical Applications for Spine Imaging: Proceedings of the Workshop held on MICCAI, 2013, Japan}, pp.~25--33, Springer, 2014.

\bibitem{kadoury2011automatic}
S.~Kadoury, H.~Labelle, and N.~Paragios, ``Automatic inference of articulated spine models in ct images using high-order markov random fields,'' {\em Medical image analysis}, vol.~15, no.~4, pp.~426--437, 2011.

\bibitem{kadoury2013spine}
S.~Kadoury, H.~Labelle, and N.~Paragios, ``Spine segmentation in medical images using manifold embeddings and higher-order mrfs,'' {\em IEEE TMI}, vol.~32, no.~7, pp.~1227--1238, 2013.

\bibitem{aslan20103d}
M.~S. Aslan, A.~Ali, {\em et~al.}, ``3d vertebrae segmentation using graph cuts with shape prior constraints,'' in {\em ICIP}, pp.~2193--2196, IEEE, 2010.

\bibitem{chu2015fully}
C.~Chu, D.~L. Belav{\`y}, {\em et~al.}, ``Fully automatic localization and segmentation of 3d vertebral bodies from ct/mr images via a learning-based method,'' {\em PloS one}, vol.~10, no.~11, p.~e0143327, 2015.

\bibitem{kelm2013spine}
B.~M. Kelm {\em et~al.}, ``Spine detection in ct and mr using iterated marginal space learning,'' {\em Medical image analysis}, vol.~17, no.~8, pp.~1283--1292, 2013.

\bibitem{cciccek20163d}
{\"O}.~{\c{C}}i{\c{c}}ek, A.~Abdulkadir, S.~S. Lienkamp, T.~Brox, and O.~Ronneberger, ``3d u-net: learning dense volumetric segmentation from sparse annotation,'' in {\em MICCAI}, pp.~424--432, Springer, 2016.

\bibitem{isensee2021nnu}
F.~Isensee, P.~F. Jaeger, S.~A. Kohl, J.~Petersen, and K.~H. Maier-Hein, ``nnu-net: a self-configuring method for deep learning-based biomedical image segmentation,'' {\em Nature methods}, vol.~18, no.~2, pp.~203--211, 2021.

\bibitem{meng2023vertebrae}
D.~Meng, E.~Boyer, {\em et~al.}, ``Vertebrae localization, segmentation and identification using a graph optimization and an anatomic consistency cycle,'' {\em Computerized Medical Imaging and Graphics}, vol.~107, p.~102235, 2023.

\bibitem{masuzawa2020automatic}
N.~Masuzawa, Y.~Kitamura, {\em et~al.}, ``Automatic segmentation, localization, and identification of vertebrae in 3d ct images using cascaded convolutional neural networks,'' in {\em MICCAI}, pp.~681--690, Springer, 2020.

\bibitem{payer2020coarse}
C.~Payer, D.~Stern, H.~Bischof, and M.~Urschler, ``Coarse to fine vertebrae localization and segmentation with spatialconfiguration-net and u-net.,'' in {\em VISIGRAPP (5: VISAPP)}, pp.~124--133, 2020.

\bibitem{gafencu2024shape}
M.-A. Gafencu, Y.~Velikova, M.~Saleh, T.~Ungi, N.~Navab, {\em et~al.}, ``Shape completion in the dark: completing vertebrae morphology from 3d ultrasound,'' {\em IJCARS}, pp.~1--9, 2024.

\bibitem{xie2020polarmask}
E.~Xie, P.~Sun, X.~Song, W.~Wang, X.~Liu, D.~Liang, C.~Shen, and P.~Luo, ``Polarmask: Single shot instance segmentation with polar representation,'' in {\em CVPR}, pp.~12193--12202, 2020.

\bibitem{pang2020spineparsenet}
S.~Pang, C.~Pang, {\em et~al.}, ``Spineparsenet: spine parsing for volumetric mr image by a two-stage segmentation framework with semantic image representation,'' {\em IEEE TMI}, vol.~40, no.~1, pp.~262--273, 2020.

\bibitem{you2022eg}
X.~You, Y.~Gu, Y.~Liu, S.~Lu, X.~Tang, and J.~Yang, ``Eg-trans3dunet: a single-staged transformer-based model for accurate vertebrae segmentation from spinal ct images,'' in {\em 2022 IEEE 19th International Symposium on Biomedical Imaging (ISBI)}, pp.~1--5, IEEE, 2022.

\bibitem{sekuboyina2017localisation}
A.~Sekuboyina, A.~Valentinitsch, J.~S. Kirschke, and B.~H. Menze, ``A localisation-segmentation approach for multi-label annotation of lumbar vertebrae using deep nets,'' {\em arXiv preprint arXiv:1703.04347}, 2017.

\bibitem{janssens2018fully}
R.~Janssens, G.~Zeng, {\em et~al.}, ``Fully automatic segmentation of lumbar vertebrae from ct images using cascaded 3d fully convolutional networks,'' in {\em ISBI}, pp.~893--897, IEEE, 2018.

\bibitem{al2018fully}
S.~M.~R. Al~Arif, K.~Knapp, and G.~Slabaugh, ``Fully automatic cervical vertebrae segmentation framework for x-ray images,'' {\em Computer methods and programs in biomedicine}, vol.~157, pp.~95--111, 2018.

\bibitem{al2018spnet}
S.~M.~R. Al~Arif, K.~Knapp, and G.~Slabaugh, ``Spnet: Shape prediction using a fully convolutional neural network,'' in {\em MICCAI}, pp.~430--439, Springer, 2018.

\bibitem{wu2023multi}
H.~Wu, J.~Zhang, Y.~Fang, Z.~Liu, N.~Wang, Z.~Cui, and D.~Shen, ``Multi-view vertebra localization and identification from ct images,'' in {\em MICCAI}, pp.~136--145, Springer, 2023.

\bibitem{chang2020multi}
H.~Chang, S.~Zhao, H.~Zheng, Y.~Chen, and S.~Li, ``Multi-vertebrae segmentation from arbitrary spine mr images under global view,'' in {\em MICCAI}, pp.~702--711, Springer, 2020.

\bibitem{dosovitskiy2020image}
A.~Dosovitskiy, L.~Beyer, A.~Kolesnikov, {\em et~al.}, ``An image is worth 16x16 words: Transformers for image recognition at scale,'' in {\em ICLR}, 2021.

\bibitem{sekuboyina2018btrfly}
A.~Sekuboyina, M.~Rempfler, J.~Kuka{\v{c}}ka, G.~Tetteh, {\em et~al.}, ``Btrfly net: Vertebrae labelling with energy-based adversarial learning of local spine prior,'' in {\em MICCAI}, pp.~649--657, Springer, 2018.

\bibitem{chen2020deep}
D.~Chen, Y.~Bai, {\em et~al.}, ``Deep reasoning networks for unsupervised pattern de-mixing with constraint reasoning,'' in {\em ICML}, pp.~1500--1509, PMLR, 2020.

\bibitem{park2022eigencontours}
W.~Park, D.~Jin, and C.-S. Kim, ``Eigencontours: Novel contour descriptors based on low-rank approximation,'' in {\em CVPR}, pp.~2667--2675, 2022.

\bibitem{bhalodia2024deepssm}
R.~Bhalodia, S.~Elhabian, J.~Adams, W.~Tao, L.~Kavan, and R.~Whitaker, ``Deepssm: A blueprint for image-to-shape deep learning models,'' {\em Medical Image Analysis}, vol.~91, p.~103034, 2024.

\bibitem{wall2003singular}
M.~E. Wall, A.~Rechtsteiner, and L.~M. Rocha, ``Singular value decomposition and principal component analysis,'' in {\em A practical approach to microarray data analysis}, pp.~91--109, Springer, 2003.

\bibitem{weisstein2005spherical}
E.~W. Weisstein, ``Spherical coordinates,'' {\em https://mathworld. wolfram. com/}, 2005.

\bibitem{bhalodia2018deepssm}
R.~Bhalodia, S.~Y. Elhabian, L.~Kavan, and R.~T. Whitaker, ``Deepssm: a deep learning framework for statistical shape modeling from raw images,'' in {\em Shape in Medical Imaging: International Workshop, ShapeMI 2018, Held in Conjunction with MICCAI 2018, Granada, Spain, September 20, 2018, Proceedings}, pp.~244--257, Springer, 2018.

\bibitem{roy2023mednext}
S.~Roy, G.~Koehler, {\em et~al.}, ``Mednext: transformer-driven scaling of convnets for medical image segmentation,'' in {\em MICCAI}, pp.~405--415, 2023.

\bibitem{valanarasu2022unext}
J.~M.~J. Valanarasu {\em et~al.}, ``Unext: Mlp-based rapid medical image segmentation network,'' in {\em MICCAI}, pp.~23--33, Springer, 2022.

\bibitem{tang2022self}
Y.~Tang, D.~Yang, W.~Li, H.~R. Roth, B.~Landman, D.~Xu, V.~Nath, and A.~Hatamizadeh, ``Self-supervised pre-training of swin transformers for 3d medical image analysis,'' in {\em CVPR}, pp.~20730--20740, 2022.

\bibitem{cheng2023sam}
J.~Cheng, J.~Ye, Z.~Deng, J.~Chen, T.~Li, H.~Wang, Y.~Su, Z.~Huang, J.~Chen, L.~Jiang, {\em et~al.}, ``Sam-med2d,'' {\em arXiv preprint arXiv:2308.16184}, 2023.

\bibitem{lorensen1998marching}
W.~E. Lorensen and H.~E. Cline, ``Marching cubes: A high resolution 3d surface construction algorithm,'' in {\em Seminal graphics: pioneering efforts that shaped the field}, pp.~347--353, 1998.

\bibitem{schroeder1998visualization}
W.~Schroeder, K.~M. Martin, and W.~E. Lorensen, {\em The visualization toolkit an object-oriented approach to 3D graphics}.
\newblock Prentice-Hall, Inc., 1998.

\bibitem{yeghiazaryan2018family}
V.~Yeghiazaryan and I.~Voiculescu, ``Family of boundary overlap metrics for the evaluation of medical image segmentation,'' {\em Journal of Medical Imaging}, vol.~5, no.~1, pp.~015006--015006, 2018.

\bibitem{kirillov2023segment}
A.~Kirillov, E.~Mintun, N.~Ravi, H.~Mao, C.~Rolland, L.~Gustafson, T.~Xiao, S.~Whitehead, A.~C. Berg, W.-Y. Lo, {\em et~al.}, ``Segment anything,'' in {\em Proceedings of the IEEE/CVF international conference on computer vision}, pp.~4015--4026, 2023.

\bibitem{ma2024segment}
J.~Ma, Y.~He, F.~Li, L.~Han, C.~You, and B.~Wang, ``Segment anything in medical images,'' {\em Nature Communications}, vol.~15, no.~1, p.~654, 2024.

\bibitem{chen2019vertebrae}
Y.~Chen, Y.~Gao, {\em et~al.}, ``Vertebrae identification and localization utilizing fully convolutional networks and a hidden markov model,'' {\em IEEE TMI}, vol.~39, no.~2, pp.~387--399, 2019.

\end{thebibliography}

\end{document}